# AUGMENTING MINDS OR AUTOMATING SKILLS? THE DIFFERENTIAL ROLE OF HUMAN CAPITAL IN GENERATIVE AI'S IMPACT ON CREATIVE TASKS

Meiling Huang[1], Ming Jin[2], and Ning Li[1]

[1]School of Economics and Management, Tsinghua University

[2]School of Management, Wuhan University of Technology

# Abstract

Generative AI is rapidly reshaping creative work, raising critical questions about its beneficiaries and societal implications. This study challenges prevailing assumptions by exploring how generative AI interacts with diverse forms of human capital in creative tasks. Through two random controlled experiments in flash fiction writing and song composition, we uncover a paradox: while AI democratizes access to creative tools, it simultaneously amplifies cognitive inequalities. Our findings reveal that AI enhances general human capital (cognitive abilities and education) by facilitating adaptability and idea integration but diminishes the value of domain-specific expertise. We introduce a novel theoretical framework that merges human capital theory with the automation-augmentation perspective, offering a nuanced understanding of human-AI collaboration. This framework elucidates how AI shifts the locus of creative advantage from specialized expertise to broader cognitive adaptability. Contrary to the notion of AI as a universal equalizer, our work highlights its potential to exacerbate disparities in skill valuation, reshaping workplace hierarchies and redefining the nature of creativity in the AI era. These insights advance theories of human capital and automation while providing actionable guidance for organizations navigating AI integration amidst workforce inequalities.

Generative AI is transforming creative industries, challenging traditional notions of human expertise and reshaping the dynamics of work. This technology offers both promise and peril: while democratizing access to creative tools, it also risks deepening cognitive and social inequalities. Scholars have highlighted generative AI's potential to augment human creativity in areas as diverse as writing, music, and visual arts (Noy & Zhang, 2023; Zhou & Lee, 2024; Nakavachara et al., 2024). Yet, others caution that such advancements may exacerbate disparities in skill valuation, favoring those who can effectively leverage AI while marginalizing others (Acemoglu et al., 2022; Doshi & Hauser, 2024; Lee & Chung, 2024; Eloundou et al., 2023). As AI evolves from a mere tool to a co-creator, understanding who benefits most from this transformation is increasingly critical—not only for individuals and organizations but also for broader societal equity.

While initial evidence suggests that generative AI can enhance creative performance, answers to this nuanced question remain elusive (Jia et al., 2023; Li et al., 2024). Some argue that AI could reduce inequality by leveling the playing field, allowing lower-performing individuals to close the performance gap (Eloundou et al., 2023; Noy & Zhang, 2023). Yet, studies also suggest that much of the observed performance gain stems from participants relying heavily on AI-generated outputs with minimal human input, resulting in automation rather than meaningful human-AI collaboration (Noy & Zhang, 2023; Doshi & Hauser, 2024). This paradox highlights the need to examine whether generative AI truly democratizes creativity or amplifies disparities by favoring those already equipped with the skills to use it effectively.

These observations point to a broader tension inherent in AI's role in the workplace. Raisch and Krakowski (2021) describe the "automation-augmentation paradox," where AI can both replace human tasks through automation and simultaneously enhance human abilities through augmentation. This framework highlights that generative AI has the potential not only to automate processes— reducing the need for human involvement in certain elements of the creative process —but also to augment human capabilities by enhancing creative and cognitive functions. These dynamics complicate our understanding of AI's broader impact, raising critical questions about its beneficiaries and its potential to reshape skill hierarchies. This distinction is critical for understanding how generative AI reshapes the dynamics of creativity and skill valuation. Will AI level the playing field, or will it widen existing gaps by privileging those with broader, more adaptable abilities? To explore these questions, we challenge conventional wisdom and offer a novel framework that redefines human-AI collaboration in creative tasks.

Specifically, we integrate the augmentation and automation framework with human capital theories to propose that generative AI has a dual effect and contrasting impact: it lowers knowledge barriers by diminishing the value of domain-specific expertise, while simultaneously increasing the importance of general human capital, such as cognitive adaptability and education (Choudhury et al., 2020; Teodoridis et al., 2019). Building on this dual effect, we develop a novel framework that differentiates between general human capital (broad, transferable skills like problem-solving and learning capacity) and specific human capital (deep, domain-specific expertise unique to particular tasks; Rietzschel et al., 2007; Teodoridis et al., 2019). This framework sheds light on how generative AI interacts unevenly with these forms of human capital, revealing its potential to both empower and marginalize. Rather than uniformly enhancing productivity, we suggest that generative AI disproportionately benefits individuals with adaptable, transferable skills, while devaluing specialized expertise. By providing this nuanced lens, our study challenges assumptions and investigates whether AI will serve as a force for democratizing creative work or as a catalyst for reinforcing inequalities in skill valuation.

To empirically test these ideas, we conducted two randomized controlled experiments examining how generative AI interacts with human capital in creative tasks. The first experiment focused on flash fiction writing, a task accessible to a broad range of individuals. Participants' general human capital (e.g., IQ and education level) and specific human capital (e.g., writing skills) were assessed, and they were randomly assigned to either work independently or collaborate with AI. To ensure ecological validity, members of the public evaluated the flash fiction, providing real-world audience judgments (Berg, 2019; Yin et al., 2024).

The second experiment extended this investigation to song lyric composition (Nelson et al., 2023), a more specialized creative domain. Participants—ranging from novices to experienced lyricists—were provided with pre-composed musical pieces and tasked with writing lyrics tailored to their assigned composition. Once completed, the songs were professionally recorded with trained singers. Public evaluations of the finished songs were again used to reflect authentic consumer responses (Berg, 2016, 2022), allowing us to capture the nuanced ways AI impacts creativity across varying levels of human capital.

Through these experiments, we reveal how generative AI's impact on creativity depends on the interplay between general and specific human capital. Our findings challenge the assumption that AI universally enhances productivity, showing that its benefits are disproportionately influenced by individuals' human capital profiles. Rather than leveling the creative playing field, AI enhances the value of general human capital—such as cognitive adaptability—while diminishing the relative importance of specialized expertise. These insights highlight the duality of AI's role: it democratizes access to creative tools but risks widening disparities based on cognitive adaptability. By examining these dynamics, our work advances understanding of human-AI collaboration, offering critical guidance for organizations and policymakers seeking to balance innovation with equity in the AI era.

## THEORETICAL DEVELOPMENT

### Generative AI and Creative Performance

Generative AI, a new generation of artificial intelligence that creates new content and solutions across various domains, has rapidly become a pivotal tool for enhancing creativity among knowledge workers (Dell'Acqua et al., 2023; Lee & Chung, 2024). A growing body of research has demonstrated its capacity to augment human performance in diverse tasks, ranging from text generation and coding assistance to complex creative endeavors such as storytelling, music composition, and visual art creation. Studies by Huang et al. (2021) and Brynjolfsson et al. (2023) have shown that generative AI can significantly increase efficiency and creativity by automating routine tasks and offering novel ideas that humans might not conceive independently.

However, while the general consensus is that generative AI improves performance, the question of who benefits most from this technology remains underexplored. Early findings, such as those from Park et al. (2023) and Noy and Zhang (2023), suggest that AI can reduce performance disparities by offering significant support to lower-performing individuals. Yet, these studies often focus on relatively simple tasks requiring minimal human input, where AI largely operates autonomously. Noy and Zhang (2023), for instance, found that AI compresses performance variance by boosting lower performers but also observed limited human-AI interaction, as many participants submitted AI-generated outputs with minimal editing. This disparity in benefit may also reflect a ceiling effect, where higher-performing individuals experience limited incremental gains relative to their lower-performing counterparts.

Consequently, these findings may not fully capture the complexities of more collaborative tasks, where deeper human-AI collaboration is required.

## A Contingent Approach: Integrating Human Capital Theory

To understand the nuanced effects of generative AI on performance, it is critical to develop a contingent approach that accounts for individual differences in human capital (Becker, 1962; Rosen, 1976). Human capital theory, widely established in organizational behavior and economics, provides a useful framework for understanding how individuals' abilities and knowledge influence their interaction with AI (Lepak & Snell, 1999; Carpenter et al., 2001; Ployhart et al., 2011).

Within this theory, human capital is typically categorized into two distinct types: general human capital and specific human capital (Coff, 1997). General human capital represents cognitive abilities and formal education that equip individuals with versatile, transferable skills (Lepak & Snell, 2002; Ritchie & Tucker-Drob, 2018). These skills enable people to quickly learn and adapt across various tasks and industries. Importantly, general human capital fosters problem-solving, critical thinking, and the ability to work with complex information (Crook et al., 2011; Ritchie & Tucker-Drob, 2018). Because these cognitive skills are broad in nature, individuals with higher levels of general human capital are capable of navigating diverse environments and performing a wide range of tasks.

On the other hand, specific human capital encompasses specialized knowledge and expertise that is narrowly focused on particular tasks, industries, or domains (Baer, 2015; Plucker & Beghetto, 2004; Tu et al., 2020). This type of capital reflects deep, technical proficiency in a specific area, allowing individuals to excel in highly specialized roles that demand extensive training and experience.

In the context of AI, the distinction between general and specific human capital becomes even more salient. While generative AI democratizes access to knowledge and facilitates the completion of tasks that once required specialized expertise, it also interacts with human capital in ways that can either amplify or diminish the relative value of these skills (Doshi & Hauser, 2024; Zhu & Zou, 2024). The contingent approach suggests that the benefits of AI are not uniformly distributed but are instead influenced by the type of human capital an individual possesses.

**The Augmentation-Automation Perspective on Generative AI and Human Capital**

Generative AI's unique features—its lack of agency and its expansive knowledge span—make it both a powerful tool and a complex variable in human AI collaboration (Rouse, 2020; Gilardi et al., 2023). These features interact differently with general and specific human capital, leading to distinct outcomes based on the type of human capital individuals possess (Pyatt & Becker, 1966; Plucker & Beghetto, 2004). The augmentation-automation framework provides a useful lens to understand this interaction, illustrating how AI either complements or substitutes human labor depending on whether individuals rely more on general or specific human capital (Raisch & Krakowski, 2021).

Generative AI's lack of agency requires human input to produce meaningful outputs, making it heavily reliant on the cognitive and evaluative capacities of users (Boussioux et al., 2024; Wang et al., 2023). This reliance means that the effectiveness of AI in creative, complex tasks is closely tied to the user's general human capital (Choudhury et al., 2020; Mariz-Perez et al., 2012). Individuals with high levels of general human capital—those equipped with cognitive versatility, critical thinking, and broad educational backgrounds—are better positioned to extract value from AI. They can assess, refine, and apply AI-generated content within complex processes such as strategic decision-making, design, and creative work (Agarwal et al., 2023; Hui et al., 2024; Rafner et al., 2023). Because these tasks require judgment, adaptation, and the integration of diverse information, AI acts as a powerful amplifier for individuals with strong general human capital. The lack of agency in AI necessitates that human oversight remains essential, meaning that those who possess broader cognitive skills will be increasingly instrumental in guiding AI towards producing meaningful, innovative outputs. This dynamic amplifies the value of general human capital, making it indispensable in an AI-augmented workplace.

At the same time, generative AI's expansive knowledge span allows it to access and apply information across a vast array of domains, fundamentally altering how tasks that traditionally relied on specific human capital are performed (Acemoglu et al., 2022; Anthony et al., 2023). In creative work, domain-specific expertise is typically acquired through years of experience, learning, and deep familiarity with the nuances of a particular field (Amabile, 2012; Lifshitz-Assaf, 2018). This expertise allows individuals to produce creative outputs informed by their specialized knowledge, which is often tied to domain-specific memory and learned associations (Baer, 2015; Bruns, 2013; Ward, 2008). However, generative AI's ability to synthesize nearly all human knowledge and understand complex connections across fields reduces the need for such narrowly focused expertise (Anthony et al., 2023; Li et al., 2024). AI's training across vast datasets allows it to not only access deep knowledge in specific areas but also combine insights from multiple domains, enabling it to perform creative tasks that were once the exclusive domain of highly specialized experts.

By integrating these two key features of generative AI—its need for human oversight and its expansive knowledge span—with the augmentation-automation framework (Raisch & Krakowski, 2021), we can better understand how AI differentially interacts with general and specific human capital (Raisch & Krakowski, 2021; Einola & Khoreva, 2023; Lee & Chung, 2024). From the augmentation perspective, generative AI enhances the capabilities of individuals with general human capital. AI tools increase cognitive and creative productivity by providing vast resources for exploration, iteration, and decision-making (Luo et al., 2021; Einola & Khoreva, 2023; Agarwal et al., 2023). Individuals with broad, adaptable skills are better equipped to harness these tools, guiding AI in ways that enhance performance on complex, non-routine creative tasks (Meincke et al., 2024; Wang et al., 2023). In this context, the demand for general human capital rises, as the role of human oversight and creative input remains critical in realizing AI's potential.

From the automation perspective, AI's expansive knowledge span enables it to perform creative tasks traditionally dominated by specific human capital, reducing the economic value of specialized knowledge (Einola & Khoreva, 2023). As AI efficiently generates creative outputs by synthesizing knowledge across domains, the demand for deep, domain-specific expertise among experts declines, while novices may find new opportunities to engage in creative processes (Dell'Acqua et al., 2023). The more AI automates creative tasks that rely on established knowledge connections, the less critical specialized human capital becomes in driving creative performance. This shift poses challenges for workers whose roles are defined by their domain-specific expertise, as AI's capacity to replicate or approximate these tasks diminishes the relative value of such expertise while simultaneously opening pathways for novices.

Building on this foundation, we now turn to the development of specific hypotheses that stem from these key mechanisms and relationships.

# HYPOTHESIS

We first posit that the use of generative AI enhances individual creativity, a baseline assumption supported by prior research showing AI's ability to boost productivity and creative output. Studies indicate that AI can augment creativity by generating new ideas, offering alternative solutions, and streamlining iteration processes in tasks like writing and consulting (Brynjolfsson et al., 2023; Doshi & Hauser, 2024). These tasks benefit from AI's strengths in synthesizing information, producing coherent narratives, and offering stylistic variations. However, in highly creative tasks—such as flash fiction and songwriting, where brevity, originality, and rapid shifts in focus are key—the impact of AI is less straightforward (Lee & Chung, 2024; Zhou & Lee, 2024). These tasks often demand novel ideas, emotional depth, and unpredictable shifts, traditionally seen as the realm of human intuition, raising questions about AI's role in enhancing creativity in such contexts.

Nevertheless, several core mechanisms suggest that AI could still improve creative performance in these highly dynamic tasks. First, AI's capacity to access and synthesize vast knowledge across genres, themes, and styles provides a wealth of inspiration, allowing users to explore novel ideas that might not be immediately apparent through human creativity alone (Marrone et al., 2024; Meincke et al., 2024). This extensive knowledge base enables individuals to combine concepts in innovative ways, potentially sparking fresh and unique creative outputs. Moreover, AI facilitates rapid iteration, allowing people to experiment with multiple creative directions (Peng et al., 2023; Nakavachara et al., 2024). This iterative process increases the likelihood of refining ideas and enhancing the final creative product. Therefore,

**Hypothesis 1.** *The use of generative AI enhances individual creativity.*

Building on the first hypothesis, which posits that generative AI enhances individual creativity, we now consider how general human capital augments this relationship. The core of this argument lies in how individuals' cognitive abilities and education level interact with AI's capabilities, particularly in creative tasks, where novelty and adaptability are key (Harvey & Berry, 2023; Doshi & Hauser, 2024; Lee & Chung, 2024).

Generative AI offers a vast array of ideas, but it lacks the ability to independently direct or refine them—relying instead on humans to guide the process (Acemoglu et al., 2022; Noy & Zhang, 2023). This is where general human capital comes into play. Individuals with high cognitive flexibility can more effectively interpret and integrate AI-generated content, drawing from a range of inputs and integrating them in unique ways (Tu et al., 2020; Meincke et al., 2024). In tasks that demand originality, those with higher education level are better equipped to navigate and synthesize AI's diverse offerings. For instance, in songwriting, an individual with broad knowledge might use AI-generated lyrics from various musical genres and styles, merging them into something fresh and innovative that goes beyond what AI alone could produce.

Additionally, the human role in providing oversight becomes critical. While AI can suggest numerous creative paths, individuals must exercise judgment to evaluate and refine these ideas (Anthony et al., 2023; Peng et al., 2023). Here, the cognitive strength associated with general human capital enables individuals to make strategic decisions about which AI-generated ideas to pursue (Boussioux et al., 2024). For example, in fiction writing, someone with high cognitive ability may discern which AI-generated plot elements will best enhance the emotional resonance or thematic complexity of the story, resulting in a more compelling final product.

Furthermore, AI's ability to draw on a vast expanse of knowledge across fields is most effectively utilized by individuals with a similarly broad base of knowledge (Jia et al., 2023; Noy

& Zhang, 2023). Those with higher levels of general human capital can connect AI-generated content to a variety of contexts, pushing creative boundaries further (Mariz-Perez et al., 2012; Dell'Acqua et al., 2023). In songwriting, for example, an individual might blend poetic, historical, and contemporary influences into their lyrics, creating something more original than either they or the AI could achieve alone.

Taken together, individuals with higher levels of general human capital are not only better at guiding AI but also at leveraging its wide-ranging capabilities to produce more innovative and impactful creative outputs (Huang et al., 2024; Rafner et al., 2023). Their ability to adapt, evaluate, and synthesize AI-generated content enhances the creative process, making the relationship between AI use and creativity particularly strong for those with greater cognitive flexibility and educational background. Therefore,

**Hypothesis 2.** *General human capital positively moderates the relationship between the use of generative AI and creativity, such that the positive relationship between AI-use and creativity will be stronger when individuals' general human capital is higher (H2a: education; H2b: IQ).*

In contrast to the synergistic interaction between AI and general human capital, generative AI may diminish the importance of specific human capital in creative tasks (Baer, 2015; Dane, 2010; Tu et al., 2020). Specific human capital, built through years of domain-specific learning and expertise, plays a vital role in producing creative outputs informed by deep knowledge (Amabile, 2012; Bruns, 2013; Teodoridis et al., 2019). However, AI's expansive knowledge span, coupled with its ability to synthesize information from diverse fields, reduces the need for narrowly focused expertise (Acemoglu & Restrepo, 2022; Eloundou et al., 2023). This shift challenges the value of specific human capital, particularly in tasks such as fiction writing and songwriting, where AI can now perform functions once requiring deep, domain-specific skills.

A key mechanism is AI's ability to automate routine elements of creative tasks. Much of specific human capital involves knowledge internalized through years of experience, such as understanding narrative structures or lyrical patterns (Zhou & Lee, 2024). For example, a professional lyricist develops an intricate understanding of lyrical structure, genre conventions, and thematic depth over time, applying these learned associations to produce high-quality compositions. However, generative AI, trained on vast knowledge corpus, can replicate these established techniques, reducing the need for domain-specific human intervention. AI's proficiency in producing creative outputs that follow conventional structures undermines the unique value that specific human capital once offered, especially in formulaic aspects of creativity.

Additionally, AI's ability to draw from a wide array of knowledge domains goes beyond the more constrained scope of specific human capital (Yin et al., 2024; Zhou & Lee, 2024). While domain-specific experts focus on the nuances of their particular field, AI can integrate diverse insights across disciplines, broadening creative possibilities (Luo et al., 2021; Lee & Chung, 2024). The fixed nature of specific human capital, often referred to as the curse of knowledge (Camerer et al., 1989), may limit flexibility in exploring ideas beyond familiar frameworks. For example, experts deeply rooted in their field may overlook novel ideas that lie outside their established knowledge base, especially when AI suggests unconventional combinations (Dane, 2010; Miller et al., 2006; Ward, 2008; Schillebeeckx et al., 2019). AI's lack of agency, requiring human oversight, further complicates this interaction, as specialists may rely too heavily on their own expertise, missing out on creative possibilities that don't align with their domain-specific knowledge (Amabile, 1985; Lawless & Kulikowich, 2006; Rietzschel et al., 2007).

Furthermore, the distinctiveness of specific skills, often developed through extensive training (Tu et al., 2020), becomes less critical when AI can replicate them at scale (Huang et al., 2024). The value of deep expertise, once a significant advantage in creative fields, is diminished when AI can produce outputs that rival or exceed the quality of those created by human experts (Doshi & Hauser, 2024; Zhou & Lee, 2024). AI's ability to emulate specific techniques and structures reduces the competitive edge of those with domain-specific skills, as the unique contributions of such expertise are no longer as essential to the creative process (Harvey & Kou, 2013; Agarwal et al., 2023).

As AI automates routine tasks, integrates diverse knowledge, and offers creative solutions beyond the confines of specific expertise, the traditional advantages of specific human capital is diminished (Puranam, 2021; Marrone et al., 2024).

**Hypothesis 3.** *Specific human capital negatively moderates the relationship between the use of generative AI and creativity, such that the positive relationship between AI-use and creativity will be weaker when the individuals' specific human capital is higher.*

## OVERVIEW OF STUDIES

We conducted two experiments to test the effects of generative AI on creativity and the moderating roles of general and specific human capital. Study 1 focused on flash-fiction writing, while Study 2 extended this investigation to a lyric-writing task, addressing the limitations of the first study and examining the interaction effects between AI use and human capital on creativity (see Figure 1 and Figure 2 for detailed experiment designs). In both studies, participants were randomly assigned to either use generative AI or complete the task independently. The AI tool was deployed via a user-friendly, dialogue-based interface built using OpenAI's API (GPT-4), allowing participants to interact seamlessly with the system (see Figure 3 for the interface of the used AI tool). By employing distinct creative contexts across the two studies, we aimed to capture a broader understanding of how AI influences creative output and how this relationship is moderated by individual differences in human capital.[1]

## EXPERIMENT 1

### Samples and Procedures

We recruited participants with a shared interest in story creation, through various channels including social media and online interest-based groups, ensuring a diverse sample comprising university students and professionals across various industries in China. Participants signed up our experiment and paid visit to our behavioral lab in schedule. 162 individuals participated in the first experiment, each compensated 30 CNY. Of the final sample, 111 (68.52%) were female, with an average age of 26.27 years ($SD = 5.62$). The majority, 154 participants (95.06%), held at least a bachelor's degree. Among them, 101 were college students, while the remaining participants worked in different sectors such as technology (8.02%) and education (6.79%).

The experiment was conducted in three stages. First, participants completed an IQ test and provided demographic information. Second, they were randomly assigned to one of two conditions: one group used generative AI (GPT-4) to compose a flash-fiction of under 500 Chinese characters, while the other group completed the task without AI assistance. Both groups were informed with basic fiction writing techniques and requirements. For AI-assist group, information about effective prompt crafting was additionally provided to ensure all participants could use AI. After the experiment, participants completed a post-experiment survey to capture their subjective perceptions during the creative process and received their compensations (see online Appendix A for measures used in survey).

---

[1] This study is part of a broader research project titled "Human Interactions with Artificial Intelligence in Organizations", which received IRB approval. All data, analysis code, output, and research materials including the full list of items are available at https://osf.io/ynhtu/?view_only=31642f7caac74082940eb1153d4e9e55. All data were analyzed using STATA MP Version 17.0.

**Measures**

 **Creativity measure.** We measured creativity using the consensual assessment technique (Amabile et al., 1996; Amabile & Pratt, 2016), following Berg's (2016, 2019) approach. We recruited an online panel of raters who evaluated the created fictions in two dimensions: novelty (ICC$_2$ = .90–.91) and usefulness (ICC$_2$ = .87–.89)[2]. Novelty was defined as the extent to which the story presented novel and distinctive ideas, reflecting originality and uniqueness. Usefulness was defined as the degree to which the story provoked thought and conveyed meaningful insights or lessons, recognizing that its value may vary based on the context of the task.

 To assess the quality of each story, we included an overall enjoyment rating from raters (ICC$_2$ = .89–.91) as an additional dimension. This measure complements the specific dimensions of novelty and usefulness, providing a broader perspective on the stories' impact. Overall enjoyment serves as a key indicator of how well the stories resonate with audiences. To ensure consistent assessments, raters participated in online training and received standardized definitions and criteria (see Appendix A). Ratings were made on a 10-point scale (1 = *Extremely low*, 10 = *Extremely high*). To mitigate potential bias perceptions against AI (Yin et al., 2024), raters indicated whether they believed each story involved generative AI (1 = *Yes*, 0 = *No*).

---

[2] When multiple groups of raters were used, the range of ICC2's is shown.

Attention checks were randomly embedded; data from two raters were excluded due to failure in these checks. Each story was evaluated by an average of 43.87 raters ($SD = 1.77$).

**General human capital.** Participants' general human capital was assessed through their educational attainment and IQ test scores (Pyatt & Becker, 1966; Crook et al., 2011; Mariz-Perez et al., 2012), both collected during the initial phase of the experiment. Participants first reported their highest level of education (1 = *junior high school and below*, 6 = *doctoral degree*). They took an 18-item version of the Raven Progressive Matrices test, which consisted of reasoning questions and had a 10-minute time limit (Sefcek et al., 2016).

**Specific human capital.** To measure the participants' specific human capital in fiction writing, we utilized self-reported assessments of their literary writing skills. This was measured with two items: "How would you rate your literary writing ability?" (1 = *Extremely poor*, 5 = *Extremely good*) and "Compared to your peers, how would you rate your literary writing ability?" (1 = *Significantly worse than most peers*, 5 = *Significantly better than most peers*). The average of these two items was used to represent participants' overall literary writing ability (Cronbach's α = .77).

**Control variables.** We controlled for several variables to ensure the robustness of our findings. First, we included demographic factors—age and gender. To account for personality traits, we controlled for openness, measured using Saucier's (1994) brief Big Five scale (8 items; e.g., "imaginative and creative"; Cronbach's α = .83). We controlled for the frequency of AI usage (0 = *never*, 5 = *daily*) because frequent AI users may be more proficient with AI tools, potentially enhancing creative outcomes due to their experience rather than the experimental conditions. Separately, we controlled for participants' mind perception of AI, measured with an adapted scale from Yam et al. (2021; 8 items; e.g., "AI can think," "AI can plan"; Cronbach's α

= .81), as individuals who perceive AI as more cognitively capable might interact differently with AI during the task, influencing their reliance on and utilization of the technology. To address potential biases related to participant motivation, we coded their motivation for participation (0 = *monetary compensation*, 1 = *other reasons such as interest in AI or fiction*). Finally, to account for potential evaluation bias toward AI (Yin et al., 2024), we controlled for the AI identification ratio, calculated as the proportion of raters who believed AI was used in creating each story.

**Results**

We employed Ordinary Least Squares (OLS) regression models to test our hypotheses. Table 1 presents the descriptive statistics and correlations among the study variables, and Table 2 provides the detailed regression results.

In supporting Hypothesis 1, AI use was positively and significantly related to novelty ($b$ = 0.403, $p$ = .035), usefulness ($b$ = 0.352, $p$ = .032), and overall impression ($b$ = 0.370, $p$ = .015).

Hypothesis 2 posits that general human capital amplifies the effect of AI use on creativity. The interaction between AI use and education was found to be positive and significant for novelty ($b$ = 0.480, $p$ = .015) and approaching significance for overall impression ($b$ = 0.309, $p$ = .064), indicating that the positive effect of AI use on creativity is stronger for individuals with higher education levels. In contrast, the moderation effect on usefulness was positive but not significant ($b$ = 0.295, $p$ = .118). Simple slope analysis revealed that, for individuals with high education, the positive effect of AI use on novelty was significant ($b$ = 0.774, $t$(149) = 3.25, $p$ = .001). Conversely, this effect was not significant for those with low education ($b$ = -0.010, $t$(149) = -0.04, $p$ = .968), as illustrated in Figure 4. A similar pattern emerged from the simple slope analysis for the usefulness and overall impression dimensions, as shown in the figures in the Appendix F. These results partially support Hypothesis 2a.

Similarly, the interaction between AI use and IQ was positive and significant for both novelty ($b$ = 0.193, $p$ = .008) and overall impression ($b$ = 0.140, $p$ = .041), and approaching significance for usefulness ($b$ = 0.106, $p$ = .060). These findings suggest that the positive effect of AI use on creativity is stronger for individuals with higher IQ levels, supporting Hypothesis 2b. Simple slope analysis confirmed a significant positive effect on novelty when IQ was high ($b$ = 0.869, $t$(149) = 3.04, $p$ = .003), while the effect was not significant when IQ was low ($b$ = -0.147, $t$(149) = -0.62, $p$ = .537), as shown in Figure 5. Similar patterns were observed for the usefulness and overall impression dimensions, with detailed results available in the online Appendix F.

Hypothesis 3 posits that specific human capital weakens the relationship between AI use and creativity. The interaction between AI use and specific human capital was negative and significant for usefulness ($b$ = -0.600, $p$ = .003) and overall impression ($b$ = -0.404, $p$ = .047), suggesting that the positive effect of AI use on creativity is diminished among individuals with higher levels of specific human capital. Although the moderation effect on novelty was negative, it was not significant ($b$ = -0.341, $p$ = .169). Further analysis revealed that for the usefulness dimension, the simple slope was positive and significantly when writing skills were low ($b$ = 0.706, $t$(149) = 3.49, $p$ = .001), but not significant when writing skills were high ($b$ = -0.065, $t$(149) = -0.31, $p$ = .758), as shown in Figure 6. Similar patterns were observed for novelty and overall impression. These findings collectively suggest partial support for Hypothesis 3.

**Supplementary Analysis**

Building on our main hypotheses, we conducted additional analyses to deepen our understanding of the effects of AI on creativity. First, we investigated whether individuals with varying levels of general and specific human capital interacted with AI differently in terms of style or mode. We conducted mean split analyses to categorize participants into high and low groups for both specific and general human capital. Specific human capital, measured by self-reported writing skills, was split at the mean score of 3.26 ($SD = 0.64$, $N_{low} = 54$, $N_{high} = 57$). Independent samples t-tests revealed no significant differences between these groups in terms of prompt length ($t(109) = 0.923$, $p = .358$) and the number of interaction rounds with the AI ($t(109) = 1.075$, $p = .285$). Participants were divided into high and low education groups based on a mean of 4.59 ($SD = 0.82$, $N_{low} = 52$, $N_{high} = 59$). T-tests showed no significant differences between high and low education groups regarding prompt length ($t(109) = -1.403$, $p = .164$) and interaction rounds ($t(109) = 0.897$, $p = .386$). Similarly, for IQ, the mean split was at 15.56 ($SD = 2.63$, $N_{low} = 62$, $N_{high} = 49$). T-tests indicated no significant differences in prompt length ($t(109) = -0.194$, $p = .846$) or interaction rounds ($t(109) = 1.05$, $p = .916$) between high and low IQ groups.

Next, considering prior research suggesting that AI use may lead to increased similarity in outputs, we employed textual analysis techniques (embedding) to assess the similarity of the creative products. Interestingly, our findings showed no significant increase in similarity among AI-assisted outputs compared to those created independently, indicating that AI use in our study did not homogenize creative work.

Third, we explored whether AI use impacted participants' cognitive perceptions of their creativity. Results revealed that using AI significantly reduced participants' psychological ownership of their creative products ($b = -1.239$, $p = .001$).

Lastly, to ensure the robustness of our main results, we conducted an *omnibus* test by including all interactions in the same regression model. The findings remained highly consistent with our initial analyses, and in several cases, the interaction effects became stronger. Together, these supplementary analyses contribute to a more comprehensive understanding of the nuanced effects of AI on creativity, supporting the robustness of our main findings. Additional details are provided in the online Appendix D.

**Experiment 1 Discussion**

Experiment 1 demonstrated that generative AI significantly enhances creativity in flash-fiction writing, positively impacting novelty, usefulness, and overall impression. Notably, individuals with higher general human capital benefited more from AI, while those with higher specific human capital experienced less benefit.

Despite testing all hypotheses, several limitations warrant consideration. First, the general nature of writing may dilute the unique impact of specific human capital, potentially explaining some insignificant moderation effects. Second, our assessment of specific human capital relied on broad self-reports of writing ability, which may not capture essential skills for novel writing, such as story development and emotional expression, leading to possible response bias. Third, participants completed tasks within a constrained timeframe in the lab, which may not reflect the extended periods typical of real-world creative processes.

To address these limitations, our second study involves a lyric-writing task with both expert and novice lyricists, allowing for a clearer operationalization of specific human capital based on lyric-writing publication history. This study also spans one week, providing participants ample time to engage deeply with the creative process, thereby enhancing ecological validity and better mimicking real-world work conditions.

## EXPERIMENT 2

**Sample and Procedures**

In Experiment 2, participants were recruited from universities, companies, and online music platforms, ensuring a diverse range of lyric-writing skills, including individuals with prior writing and publication experience. To incentivize participation and engagement, each participant was promised a professionally recorded song composition based on their own lyric creation, in addition to receiving 100 CNY upon completing all stages of the experiment.

The participants were tasked with writing song lyrics, a key component of a song alongside vocal melodies and instrumental accompaniments. To support this task, we provided each participant with both a vocal melody and instrumental accompaniment. We prepared ten royalty-free accompaniment tracks in various styles, and two professional composers created vocal melodies for five tracks each, resulting in ten complete song demos. Each demo consisted of an accompaniment track paired with a vocal melody performed using "la-la-la" syllables (see the online Appendix B for delivered materials).

After registration and an online IQ test ($N_{start}$ = 685), participants were randomly assigned to either an AI-assisted group or a control group that composed lyrics without AI support. Both groups received basic instructions on lyric-writing techniques, while the AI-assisted group received additional guidance on using generative AI ($N_{information}$ = 611). Participants were assigned a demo file along with a simplified musical score of the vocal melody, which included annotations for lyric breaks and suggested word counts. This design aimed to engage participants effectively, regardless of their experience level.

The experiment was conducted online, allowing participants one week to complete the task at their own pace, closely mimicking typical lyrics-writing processes. Throughout this period, participants could listen to the provided melody and refer to the musical score as they composed their lyrics, facilitating a structured and supportive creative environment. After completing their initial assignment, participants were encouraged to write lyrics for the remaining nine demos. They then submitted their lyrics and completed a follow-up questionnaire about their experiences ($N_{submission}$ = 348).

Sample attrition occurred primarily during the lyrics-creation stage, which had a dropout rate of 43.04% despite efforts to simplify the task. Among the submissions, 329 works from 299 participants were deemed suitable for song recording by professional producers and singers. Of these participants, 289 composed lyrics for only one song (96.66%). The final sample of 299 participants had an average age of 24.10 years ($SD$ = 6.06); 191 (63.88%) were female, and 279 (93.31%) held a bachelor's degree or higher.

**Measures**

**Creativity measure.** In Experiment 2, we employed a dual-method approach to assess the creativity of the composed lyrics, ensuring a comprehensive evaluation that captures both the intrinsic qualities of the lyrics and their reception within a musical context. First, we recruited an online panel of raters to evaluate the lyrics independently of their musical performance, thereby minimizing potential influences from accompanying melodies and arrangements. These raters rated each lyric on three dimensions: novelty ($ICC_2$ = .35–.52), defined as the originality and unique expression within the lyrics, including innovative rhetorical techniques and perspectives; usefulness (emotional expression; $ICC_2$ = .26–.41), referring to the extent to which the emotional content resonates with and engages the audience; and overall impression ($ICC_2$ = .42–.59), which pertains to the overall quality and impact of the lyrics themselves. To ensure consistency, raters participated in online training sessions and were provided with standardized definitions and criteria for each dimension (see online Appendix C). Ratings were conducted using a 10-point scale (1 = *Extremely low*, 10 = *Extremely high*), with attention checks embedded throughout the process, resulting in the exclusion of data from two raters. Each lyric was evaluated by an average of 5.12 raters ($SD = 0.71$).

Second, to evaluate the creativity of the lyrics within a musical context, we composed the written lyrics into complete songs. Ten members from university choirs, all with systematic and professional training in singing, performed the composed lyrics. A professional music producer then finalized these performances into complete song recordings. An online panel of raters assessed the lyrics on the same two dimensions of novelty ($ICC_2 = .85–.87$) and usefulness (emotional appeal; $ICC_2 = .78–.81$), maintaining the same criteria. Additionally, raters evaluated their overall impression of the songs based on their liking as listeners ($ICC_2 = .83–.85$), reflecting the audience's reception of the complete musical piece. This second rating phase allowed us to consider the fit between the lyrics and their musical execution. Raters received training and detailed definitions to ensure consistent evaluations. Attention checks were included, leading to the exclusion of five raters who failed these checks and two raters whose correlation coefficients with the average scores were below .3. Each song was evaluated by an average of 28.33 raters ($SD = 1.61$).

**General human capital.** Consistent with Experiment 1, participants' general human capital was assessed through their education level and IQ test scores using the same measurement (Mariz-Perez et al., 2012; Ployhart et al., 2011; Sefcek et al., 2016).

**Specific human capital.** We measured participants' specific human capital in the field of lyrics writing using an objective measurement. One single indicator was employed to measure participants' previous experience with publishing or showcasing their lyrics: "Have you ever published or presented your lyrics?" The responses were distributed as follows—1 (No) accounted for 64.44%, 2 (Yes, but only within a small range) for 29.79%, 3 (Yes, on a public platform) for 4.86%, and 4 (My work has received awards or recognition) for 0.91%. Compared to self-assessed writing ability, participants' experience in publishing or showcasing their lyrics provides a more objective measure and directly reflects their accumulated experience in this field.

**Control variables.** Consistent with Experiment 1, we controlled for participants' age, gender, openness, frequency of AI usage, motivation for participating in the experiment, mind perception of AI, and the AI identification ratio by raters. All the measurements stayed the same with Experiment 1. Additionally, we included fixed effects for the selected demo in our regression analysis.

## Results

We employed Ordinary Least Squares (OLS) regression models to test our hypotheses. Descriptive statistics and correlations are presented in Table 3, and regression results are shown in Tables 4–8.

The results provide limited support for hypothesis 1. When evaluating lyrics alone, AI use did not significantly predict creativity. In contrast, when assessing complete songs, AI use showed positive coefficients for novelty ($b = 0.133$, $p = .061$) and usefulness ($b = 0.108$, $p = .075$), though these effects did not reach statistical significance.

Hypothesis 2a proposed that education would moderate the relationship between AI use and creativity, with stronger effects for individuals with higher education levels. For lyrics ratings, the interaction between AI use and education was significant across all creativity dimensions: novelty ($b = 0.407$, $p = .008$), usefulness ($b = 0.341$, $p = .021$), and overall impression ($b = 0.489$, $p = .002$). Simple slope analyses revealed that for individuals with high education, AI use positively influenced novelty ($b = 0.389$, $t(308) = 2.49$, $p = .013$; see Figure 7), usefulness ($b = 0.333$, $t(308) = 2.22$, $p = .027$), and overall impression ($b = 0.461$, $t(308) = 2.93$, $p = .004$). For those with low education, the effects of AI use were negative but not significant across all dimensions. In the context of complete songs, the interaction effects between AI use and education were positive but not statistically significant, although the direction remained consistent with our hypothesis. These results indicate that the positive effect of AI use on creativity is stronger among more educated individuals when evaluating lyrics alone, providing partial support for Hypothesis 2a.

Hypothesis 2b suggested that IQ would moderate the relationship between AI use and creativity, with stronger effects for individuals with higher IQ scores. The interactions were not statistically significant for any creativity dimensions in either lyrics ratings or complete songs. Although the coefficients were in the expected direction, we do not find support for Hypothesis 2b.

Hypothesis 3 posited that specific human capital, measured by prior lyrics publication experience, would negatively moderate the relationship between AI use and creativity. The results support this hypothesis across both lyrics ratings and complete songs. For lyrics ratings, the interaction between AI use and specific human capital was significantly negative for novelty ($b = -0.391$, $p = .017$) and was approaching significance for usefulness ($b = -0.327$, $p = .056$) and overall impression ($b = -0.283$, $p = .091$). Similarly, for complete songs, the interaction was significantly negative for novelty ($b = -0.221$, $p = .029$), usefulness ($b = -0.195$, $p = .017$), and overall impression ($b = -0.209$, $p = .014$). Simple slope analyses showed that for individuals with low specific human capital, AI use positively influenced creativity—significantly for novelty ($b = 0.264$, $t(308) = 2.76$, $p = .006$; See Figure 8), usefulness ($b = 0.223$, $t(308) = 2.75$, $p = .006$), and overall impression ($b = 0.211$, $t(308) = 2.568$, $p = .011$) in the context of complete songs.

Conversely, for those with high specific human capital, the effects of AI use on creativity were negative but not statistically significant across all dimensions. These findings indicate that the positive effects of AI use on creativity are weaker for individuals with higher levels of specific human capital, supporting Hypothesis 3.

**Supplementary Analysis**

We also conducted several supplementary analyses similar to first study. First, we categorize participants into high and low groups for specific human capital by identifying whether they have any lyrics publication previously ($N_{low}$ = 131, $N_{high}$ = 66). Independent samples t-tests revealed no significant differences between these groups in terms of prompt length ($t$(195) = 1.539, $p$ = .126). However, experts interact with AI significantly less than novice regarding the number of interaction rounds with the AI ($t$(195) = 2.207, $p$ = .029). Similar to Experiment 1, we conducted mean split analyses to categorize participants into high and low groups for general human capital. For general human capital, participants were divided into high and low education groups based on a mean of 4.31 ($SD$ = 0.71, $N_{low}$ = 132, $N_{high}$ = 65). T-tests showed individual with higher education interact significantly more with AI than their counterparts with lower education level regarding both prompt length ($t$(195) = -3.189, $p$ = .002) and interaction rounds ($t$(195) = -2.206, $p$ = .029). Similarly, for IQ, the mean split was at 14.47 ($SD$ = 3.04, $N_{low}$ = 97, $N_{high}$ = 100). T-tests indicated individuals with higher IQ interact with AI approaching significantly more than their counterparts in prompt length ($t$(195) = -1.925, $p$ = .056) but no significant difference in interaction rounds ($t$(195) = -1.222, $p$ = .223) between high and low IQ groups.

Second, we again assessed the similarity of the creative products using textual analysis techniques (embedding). Contrast to Experiment 1, the results showed increased similarity among AI-assisted outputs compared to those created without AI, remaining confusion about whether AI use homogenize the creative work (Cosine: $b = 0.014$, $p = .008$; L2 distance: $b = -0.014$, $p = .007$).

Third, we investigated whether the use of AI affected participants cognitive perceptions. Results showed using AI significantly reduced participants' psychological ownership to their creativity product ($b = -1.124$, $p = .000$). Different from Experiment 1, AI use in the second study also increases participants' creative self-efficacy ($b = 0.284$, $p = .014$).

Consistent with Experiment 1, to ensure the robustness of our main results, we conducted an omnibus test by including all interactions in the same regression model. The findings remained highly consistent with our initial analyses. Details were shown in online Appendix E.

## Experiment 2 Discussion

Unlike in Experiment 1, working with generative AI in Experiment 2 did not significantly improve creativity. The moderating role of education, which was evident in the lyric-writing task, diminished when evaluating the full songs. Additionally, IQ did not significantly moderate the relationship between AI use and creativity in either case. However, specific human capital consistently moderated the AI-creativity relationship negatively across both lyrics and songs, indicating that individuals with higher domain-specific expertise benefited less from AI assistance. These differences could stem from the complexity and specificity of the songwriting task. Songwriting, as a more specialized creative domain, may reduce the impact of general human capital while amplifying the importance of specific expertise. The clearer distinction between experts and novices in Experiment 2, based on prior lyrics publication, likely intensified the negative moderation effect of specific human capital.

## GENERAL DISCUSSION

This research examined how generative AI interacts with different forms of human capital to influence creativity. Across two studies—flash fiction writing and songwriting—we explored how AI affects creativity and how general human capital (education and IQ) and specific human capital (domain-specific expertise) moderate these effects. The results reveal that AI significantly enhances creativity, especially for individuals with higher levels of general human capital. However, specific human capital consistently moderated this relationship negatively, indicating that individuals with greater domain expertise benefited less from AI assistance. These findings suggest that AI's impact on creativity is uneven, favoring those with broader cognitive skills while offering diminished advantages for those with specialized knowledge.

**Theoretical Implications**

Our study makes several important theoretical contributions. First, it challenges the notion that generative AI uniformly enhances productivity and reduces performance disparities among individuals (Noy & Zhang, 2023). Contrary to prior research on human-AI interactions, which suggests that domain experts may benefit more from AI due to their ability to effectively utilize predictive algorithms (e.g., Agrawal et al., 2019; Huang et al., 2024), our findings reveal that generative AI—unlike traditional predictive AI—can actually reduce the competitive edge of domain experts. By democratizing access to knowledge, generative AI breaks down traditional barriers, allowing individuals without specific expertise to perform tasks previously reserved for specialists (Anthony et al., 2023; Brynjolfsson et al., 2023; Wang et al., 2023). This shift underscores a fundamental change in the dynamics of knowledge work, where general cognitive skills become more valuable than specialized knowledge.

Second, by integrating human capital theory with the context of generative AI, we develop a novel framework that explains how different forms of human capital interact with AI technologies. Our findings illustrate that augmentation and automation coexist in the AI-human collaboration landscape and that their relative influence depends on the type of human capital individuals possess. Specifically, generative AI augments the capabilities of those with high general human capital by enhancing their ability to process and integrate vast amounts of information creatively. In contrast, it automates tasks traditionally reliant on specific human capital, thereby reducing the unique value of specialized expertise. This framework advances human capital theory by demonstrating that the value of different skill types is reshaped in the presence of generative AI. It explains why experts may not benefit more from AI: the breaking of knowledge barriers by AI diminishes the exclusivity of their expertise. Additionally, experts may engage less with AI tools due to factors such as AI aversion or overreliance on their own knowledge, limiting their ability to leverage AI effectively (Doshi & Hauser, 2024; Yin et al., 2024). Our research thus highlights the need to reconsider how specific and general human capital are valued in future work.

Third, our study uncovers nuanced insights into the limitations of generative AI. In Experiment 2, we did not observe significant main effects of AI use on creativity when evaluating lyrics alone, diverging from previous studies that reported consistent positive effects (Jia et al., 2023; Noy & Zhang, 2023). This suggests that AI's effectiveness may depend on the nature of the task. For instance, songwriting relies less on writing skills—a strength of generative AI—and more on idea generation and emotional expression, which may not be as readily enhanced by AI assistance.

Furthermore, we did not consistently observe the hypothesized homophily effect, which suggests that AI use leads to increased similarity in outputs (Wang et al., 2023; Anthony et al., 2023). While some studies argue that AI can homogenize creative products due to reliance on common algorithms, our findings indicate that this effect is not consistent and may vary depending on the type of task and the level of human-AI interaction.

Finally, our exploration into participants' perceptions revealed that AI use could impact intrinsic motivation. Some participants reported reduced feelings of ownership over their creative work when using AI, potentially diminishing intrinsic motivation (Amabile & Pratt, 2016). However, AI assistance also appeared to boost self-efficacy in creative domains, encouraging individuals to engage in tasks they might have otherwise avoided due to perceived skill gaps (Anthony et al., 2023; Noy & Zhang, 2023). These contrasting effects suggest that AI's influence on motivation is complex and warrants further investigation.

**Practical Implications**

Our findings have important practical implications for organizations navigating the integration of AI in creative and knowledge-based work. As AI becomes more prevalent across industries, understanding how different forms of human capital interact with AI can inform talent acquisition, workforce development, and task allocation strategies (Dell'Acqua et al., 2023; Frank et al., 2019; Paudel, 2024). Organizations should recognize the increasing value of general human capital—skills such as critical thinking, problem-solving, and adaptability—in an AI-enhanced workplace. Prioritizing these skills in hiring and training programs can enhance employees' ability to collaborate effectively with AI technologies.

Companies can invest in developing general cognitive skills through targeted training initiatives, thereby maximizing the benefits of AI integration. At the same time, industries heavily reliant on domain-specific expertise may need to reconsider the role of such knowledge in an AI-driven economy (Allen & Choudhury, 2022; Brynjolfsson et al., 2023). Our findings suggest that AI's capacity to automate specialized tasks could reduce the competitive advantage of individuals with narrowly focused expertise. Organizations might therefore shift toward fostering interdisciplinary skills and encouraging employees to develop broader competencies.

From a societal perspective, policymakers and educators should emphasize broad-based educational programs that cultivate general cognitive abilities, ensuring that individuals are equipped to thrive alongside AI technologies (Furman & Seamans, 2019; Frank et al., 2019). Strategies to mitigate potential inequalities exacerbated by differences in human capital are essential, promoting inclusive access to skills development opportunities.

**Limitations and Future Directions**

Despite the contributions of our research, several limitations warrant acknowledgment and present avenues for future investigation. First, in Experiment 1, the distinction between high and low specific human capital may not have been salient due to reliance on broad self-assessments of writing ability. Experiment 2 addressed this limitation by using prior lyrics publication as a clearer indicator of specific human capital, resulting in more consistent results. Future research should employ precise and validated measures of specific human capital to better capture its nuances across different creative domains.

Second, some inconsistencies between Experiment 1 and Experiment 2, particularly regarding the main effects of AI use and the moderating role of general human capital (e.g., IQ), suggest that the impact of AI may vary across tasks. Songwriting may rely less on writing skills—a strength of generative AI—and more on idea generation and emotional expression, areas where AI assistance may be less effective. Future research should explore a range of creative tasks to determine the conditions under which AI enhances or diminishes creativity.

Third, our measures of general human capital—IQ tests based on Raven's Progressive

Matrices and education level—focus on logic, reasoning, and cognitive skills that may not directly translate to artistic creativity (Ritchie & Tucker-Drob, 2018). This may explain why IQ and education did not predict performance directly in our studies. Additionally, cultural factors, such as the emphasis on logic and mathematics in Chinese education, may limit the applicability of these measures to creative tasks. Future studies should consider alternative measures of general human capital that capture a broader range of cognitive abilities relevant to creativity.

Finally, our research focused on creative tasks involving writing and lyric creation. It remains to be seen whether similar patterns emerge in tasks involving different cognitive demands, such as logical reasoning, coding, or analytical problem-solving. Investigating the interaction of AI and human capital in diverse domains would enhance the generalizability of our theoretical framework and inform AI integration strategies across various industries.

## Conclusion

In conclusion, our study provides valuable insights into the complex interplay between generative AI and human capital in creative work. By demonstrating that AI does not uniformly enhance productivity and that its benefits are contingent on the type of human capital individuals possess, we contribute to a more nuanced understanding of AI's role in the modern workplace. These findings have significant implications for theory, practice, and future research, highlighting the need to reconsider how we value and develop human skills in an era increasingly shaped by AI technologies.

**Table 1 Means, SDs, and Correlation of Studied Variables (Experiment 1)**

| | VARIABLES | Mean | SD | 1 | 2 | 3 | 4 | 5 | 6 | 7 |
|---|---|---|---|---|---|---|---|---|---|---|
| 1 | Novelty | 4.879 | 1.036 | - | | | | | | |
| 2 | Usefulness | 4.799 | 0.940 | 0.744••• | - | | | | | |
| 3 | Overall Impression | 4.912 | 0.918 | 0.870••• | 0.881••• | - | | | | |
| 4 | Education | 4.586 | 0.816 | -0.086 | -0.086 | -0.155• | - | | | |
| 5 | IQ | 15.556 | 2.626 | -0.093 | -0.049 | -0.116 | 0.206•• | - | | |
| 6 | Specific Human Capital | 3.269 | 0.643 | 0.008 | 0.040 | 0.016 | 0.065 | 0.031 | - | |
| 7 | Age | 26.272 | 5.622 | -0.117 | -0.124 | -0.186• | 0.305••• | -0.144 | 0.024 | - |
| 8 | Gender | 1.685 | 0.466 | -0.048 | -0.094 | -0.041 | 0.031 | -0.100 | 0.097 | -0.022 |
| 9 | Openness | 3.854 | 0.607 | 0.053 | 0.072 | 0.042 | 0.098 | 0.008 | 0.364••• | 0.019 |
| 10 | AI Use Frequency | 2.204 | 1.458 | -0.104 | -0.114 | -0.132 | 0.191• | 0.329••• | -0.016 | -0.135 |
| 11 | Purpose for Experiment | 0.056 | 0.230 | 0.180• | 0.219•• | 0.189• | 0.024 | -0.124 | -0.207•• | -0.021 |
| 12 | Mind Perception | 3.983 | 0.827 | -0.176• | -0.096 | -0.125 | -0.019 | -0.003 | -0.010 | -0.036 |
| 13 | AI Identification Ratio | 0.369 | 0.134 | -0.262••• | -0.404••• | -0.441••• | -0.004 | -0.049 | -0.065 | 0.144 |

*Continued*

| | VARIABLES | 8 | 9 | 10 | 11 | 12 | 13 |
|---|---|---|---|---|---|---|---|
| 8 | Gender | - | | | | | |
| 9 | Openness | -0.032 | - | | | | |
| 10 | AI Use Frequency | -0.042 | 0.041 | - | | | |
| 11 | Purpose for Experiment | -0.242•• | -0.103 | 0.059 | - | | |
| 12 | Mind Perception | 0.010 | 0.050 | -0.020 | -0.011 | - | |
| 13 | AI Identification Ratio | -0.044 | -0.138 | -0.061 | -0.122 | 0.101 | - |

*Notes. Female = 2, Male = 1. All p values in this table are two-tailed.*
*••• p<0.001, •• p<0.01, • p<0.05.*

## Table 2 Regression Results (Experiment 1)

| VARIABLES | Model 1 | | | Model 2a | | | Model 2b | | | Model 3 | | |
|---|---|---|---|---|---|---|---|---|---|---|---|---|
| | Novelty | Usefulness | Overall Impression | Novelty | Usefulness | Overall Impression | Novelty | Usefulness | Overall Impression | Novelty | Usefulness | Overall Impression |
| AI Use | 0.403• | 0.352• | 0.370• | 0.382• | 0.339• | 0.357• | 0.361• | 0.329• | 0.340• | 0.386• | 0.321• | 0.349• |
| | (0.19) | (0.16) | (0.15) | (0.19) | (0.16) | (0.15) | (0.18) | (0.16) | (0.15) | (0.19) | (0.16) | (0.15) |
| AI Use × Education | | | | 0.480• | 0.295 | 0.309+ | | | | | | |
| | | | | (0.20) | (0.19) | (0.17) | | | | | | |
| AI Use × IQ | | | | | | | 0.193•• | 0.106+ | 0.140• | | | |
| | | | | | | | (0.07) | (0.06) | (0.07) | | | |
| AI Use × Specific Human Capital | | | | | | | | | | -0.341 | -0.600•• | -0.404• |
| | | | | | | | | | | (0.25) | (0.20) | (0.20) |
| Education | -0.040 | -0.048 | -0.087 | -0.367• | -0.249 | -0.298• | -0.036 | -0.046 | -0.084 | -0.043 | -0.053 | -0.091 |
| | (0.11) | (0.09) | (0.08) | (0.17) | (0.17) | (0.14) | (0.10) | (0.09) | (0.08) | (0.11) | (0.08) | (0.08) |
| IQ | -0.019 | 0.000 | -0.025 | -0.020 | -0.000 | -0.025 | -0.164• | -0.079 | -0.130• | -0.021 | -0.003 | -0.027 |
| | (0.04) | (0.03) | (0.03) | (0.04) | (0.03) | (0.03) | (0.07) | (0.05) | (0.06) | (0.04) | (0.03) | (0.03) |
| Specific Human Capital | -0.024 | 0.033 | -0.005 | -0.010 | 0.042 | 0.004 | -0.008 | 0.042 | 0.007 | 0.214 | 0.451•• | 0.276 |
| | (0.13) | (0.11) | (0.11) | (0.14) | (0.11) | (0.11) | (0.13) | (0.11) | (0.11) | (0.22) | (0.17) | (0.18) |
| Age | -0.016 | -0.010 | -0.019 | -0.018 | -0.012 | -0.020+ | -0.018 | -0.012 | -0.021+ | -0.018 | -0.014 | -0.021+ |
| | (0.01) | (0.01) | (0.01) | (0.01) | (0.01) | (0.01) | (0.01) | (0.01) | (0.01) | (0.01) | (0.01) | (0.01) |
| Gender | -0.039 | -0.131 | -0.056 | -0.015 | -0.116 | -0.041 | -0.057 | -0.141 | -0.069 | -0.001 | -0.064 | -0.011 |
| | (0.17) | (0.15) | (0.13) | (0.17) | (0.15) | (0.13) | (0.17) | (0.15) | (0.13) | (0.17) | (0.14) | (0.13) |
| openness | 0.102 | 0.065 | 0.029 | 0.072 | 0.047 | 0.010 | 0.056 | 0.040 | -0.005 | 0.118 | 0.094 | 0.048 |
| | (0.12) | (0.11) | (0.10) | (0.12) | (0.11) | (0.10) | (0.12) | (0.11) | (0.10) | (0.12) | (0.11) | (0.10) |
| AI Use Frequency | -0.080 | -0.092+ | -0.085+ | -0.093 | -0.100+ | -0.094• | -0.066 | -0.085 | -0.075 | -0.073 | -0.080 | -0.077 |
| | (0.06) | (0.05) | (0.05) | (0.06) | (0.05) | (0.04) | (0.06) | (0.05) | (0.05) | (0.06) | (0.05) | (0.05) |
| Purpose for Experiment | 0.703+ | 0.735• | 0.549+ | 0.670+ | 0.714• | 0.528+ | 0.457 | 0.600• | 0.371 | 0.753+ | 0.822• | 0.608+ |
| | (0.38) | (0.34) | (0.32) | (0.39) | (0.34) | (0.32) | (0.35) | (0.32) | (0.29) | (0.41) | (0.32) | (0.33) |
| Mind Perception | -0.211• | -0.080 | -0.109 | -0.202• | -0.074 | -0.103 | -0.183• | -0.064 | -0.089 | -0.222• | -0.100 | -0.123+ |
| | (0.08) | (0.07) | (0.07) | (0.09) | (0.07) | (0.07) | (0.09) | (0.07) | (0.07) | (0.09) | (0.07) | (0.07) |
| AI Identification Ratio | -2.093•• | -2.966••• | -3.185••• | -2.131••• | -2.989••• | -3.209••• | -2.078••• | -2.957••• | -3.174••• | -2.048•• | -2.886••• | -3.131••• |
| | (0.63) | (0.57) | (0.56) | (0.61) | (0.57) | (0.55) | (0.59) | (0.56) | (0.52) | (0.64) | (0.57) | (0.56) |
| Constant | 6.996••• | 6.482••• | 7.704••• | 6.945••• | 6.344••• | 7.388••• | 6.810••• | 6.549••• | 7.403••• | 6.919••• | 6.593••• | 7.690••• |
| | (1.12) | (0.98) | (0.94) | (1.10) | (0.99) | (0.93) | (0.91) | (0.81) | (0.75) | (1.10) | (0.90) | (0.88) |
| Observations | 162 | 162 | 162 | 162 | 162 | 162 | 162 | 162 | 162 | 162 | 162 | 162 |
| R-squared | 0.178 | 0.263 | 0.315 | 0.207 | 0.276 | 0.330 | 0.223 | 0.279 | 0.345 | 0.187 | 0.295 | 0.330 |

*Notes.* Robust standard errors in parentheses. All p values in this table are two-tailed. Education was centered in Model 2a; IQ was centered in Model 2b; Specific Human Capital was centered in Model 3.
••• p<0.001, •• p<0.01, • p<0.05, + p<0.1

**Table 3 Means, SDs, and Correlations of the Studied Variables (Experiment 2)**

| | VARIABLES | Mean | SD | 1 | 2 | 3 | 4 | 5 | 6 | 7 | 8 |
|---|---|---|---|---|---|---|---|---|---|---|---|
| 1 | Novelty_L | 5.012 | 1.015 | - | | | | | | | |
| 2 | Usefulness_L | 5.374 | 0.950 | 0.808••• | - | | | | | | |
| 3 | Overall Impression_L | 5.366 | 1.026 | 0.882••• | 0.846••• | - | | | | | |
| 4 | Novelty_S | 4.836 | 0.744 | 0.618••• | 0.498••• | 0.633••• | - | | | | |
| 5 | Usefulness_S | 5.143 | 0.636 | 0.469••• | 0.479••• | 0.526••• | 0.792••• | - | | | |
| 6 | Overall Impression_S | 4.944 | 0.681 | 0.444••• | 0.417••• | 0.513••• | 0.797••• | 0.895••• | - | | |
| 7 | Education | 4.307 | 0.711 | 0.119• | 0.151•• | 0.134• | 0.101 | 0.090 | 0.052 | - | |
| 8 | IQ | 14.465 | 3.038 | 0.130• | 0.096 | 0.101 | 0.072 | 0.059 | 0.051 | 0.221••• | - |
| 9 | Specific Human Capital | 1.422 | 0.630 | 0.063 | 0.074 | 0.098 | 0.078 | 0.074 | 0.092 | -0.100 | -0.041 |
| 10 | AI Identification Raio_L | 0.626 | 0.231 | -0.231• | -0.273••• | -0.188••• | -0.057 | -0.026 | -0.009 | -0.055 | -0.011 |
| 11 | AI Identification Ratio_S | 0.569 | 0.114 | -0.134• | -0.223••• | -0.106 | -0.231••• | -0.337••• | -0.299••• | -0.090 | -0.075 |
| 12 | Age | 24.617 | 6.267 | 0.111• | 0.125• | 0.085 | 0.109• | 0.047 | 0.008 | 0.097 | 0.112• |
| 13 | Gender | 1.623 | 0.485 | 0.057 | 0.085 | 0.079 | 0.033 | 0.059 | 0.021 | 0.080 | 0.003 |
| 14 | Openness | 3.961 | 0.571 | -0.063 | -0.048 | -0.075 | -0.068 | -0.040 | -0.050 | 0.067 | -0.079 |
| 15 | AI Use Frequency | 2.900 | 1.651 | -0.040 | -0.004 | -0.014 | -0.001 | -0.027 | -0.023 | 0.213••• | 0.063 |
| 16 | Purpose for Experiment | 0.049 | 0.215 | 0.018 | -0.021 | -0.011 | 0.082 | 0.092 | 0.101 | 0.002 | 0.045 |
| 17 | Mind Perception | 3.856 | 0.861 | -0.091 | -0.090 | -0.096 | -0.034 | -0.091 | -0.084 | -0.083 | -0.050 |

*Continued*

| | VARIABLES | 9 | 10 | 11 | 12 | 13 | 14 | 15 | 16 | 17 |
|---|---|---|---|---|---|---|---|---|---|---|
| 9 | Specific Human Capital | - | | | | | | | | |
| 10 | AI Identification Raio_L | -0.028 | - | | | | | | | |
| 11 | AI Identification Ratio_S | -0.039 | 0.359••• | - | | | | | | |
| 12 | Age | 0.219••• | -0.069 | -0.080 | - | | | | | |
| 13 | Gender | -0.245••• | -0.038 | -0.120• | -0.227••• | - | | | | |
| 14 | Openness | 0.078 | -0.056 | -0.022 | 0.076 | 0.022 | - | | | |
| 15 | AI Use Frequency | -0.117• | 0.013 | 0.0480 | -0.190••• | 0.097 | 0.132• | - | | |
| 16 | Purpose for Experiment | -0.107 | -0.005 | -0.049 | -0.029 | 0.001 | -0.229••• | -0.038 | - | |
| 17 | Mind Perception | 0.115• | 0.001 | 0.164••• | 0.060 | -0.116• | 0.192••• | 0.042 | -0.054 | - |

*Notes. Female = 2, Male = 1. All p values in this table are two-tailed. "_L" refers to rating scores on lyrics only; "_S" refers to rating scores with songs hearing.*
*••• $p<0.001$, •• $p<0.01$, • $p<0.05$.*

**Table 4 Regression Results of AI Use on Creativity Measured by Lyrics (Experiment 2)**

| VARIABLES | (1) Novelty_L | (2) Novelty_L | (3) Usefulness_L | (4) Usefulness_L | (5) Overall Impression_L | (6) Overall Impression_L |
|---|---|---|---|---|---|---|
| AI Use |  | 0.087 |  | 0.079 |  | 0.097 |
|  |  | (0.11) |  | (0.11) |  | (0.12) |
| Education | 0.159$^{\bullet}$ | 0.156$^{+}$ | 0.186$^{\bullet}$ | 0.183$^{\bullet}$ | 0.201$^{\bullet}$ | 0.198$^{\bullet}$ |
|  | (0.08) | (0.08) | (0.08) | (0.08) | (0.08) | (0.08) |
| IQ | 0.029 | 0.030 | 0.017 | 0.017 | 0.015 | 0.015 |
|  | (0.02) | (0.02) | (0.02) | (0.02) | (0.02) | (0.02) |
| Specific Human Capital | 0.115 | 0.109 | 0.116 | 0.110 | 0.190$^{\bullet}$ | 0.184$^{\bullet}$ |
|  | (0.08) | (0.08) | (0.08) | (0.08) | (0.08) | (0.08) |
| Age | 0.013 | 0.013 | 0.016$^{+}$ | 0.016 | 0.010 | 0.009 |
|  | (0.01) | (0.01) | (0.01) | (0.01) | (0.01) | (0.01) |
| Gender | 0.123 | 0.117 | 0.192$^{+}$ | 0.186$^{+}$ | 0.193 | 0.186 |
|  | (0.11) | (0.11) | (0.11) | (0.11) | (0.12) | (0.12) |
| Openness | -0.093 | -0.085 | -0.119 | -0.111 | -0.158$^{+}$ | -0.148 |
|  | (0.10) | (0.09) | (0.09) | (0.09) | (0.09) | (0.09) |
| AI Use frequency | -0.028 | -0.028 | -0.001 | -0.002 | -0.012 | -0.012 |
|  | (0.04) | (0.04) | (0.04) | (0.04) | (0.03) | (0.03) |
| Purpose for Experiment | 0.076 | 0.080 | -0.114 | -0.111 | -0.017 | -0.013 |
|  | (0.24) | (0.24) | (0.23) | (0.23) | (0.25) | (0.25) |
| Mind Perception | -0.078 | -0.078 | -0.076 | -0.076 | -0.094 | -0.094 |
|  | (0.06) | (0.06) | (0.06) | (0.06) | (0.06) | (0.06) |
| AI Identification Ratio_L | -1.382$^{\bullet\bullet\bullet}$ | -1.404$^{\bullet\bullet\bullet}$ | -1.378$^{\bullet\bullet\bullet}$ | -1.398$^{\bullet\bullet\bullet}$ | -1.233$^{\bullet\bullet\bullet}$ | -1.259$^{\bullet\bullet\bullet}$ |
|  | (0.22) | (0.22) | (0.21) | (0.21) | (0.23) | (0.23) |
| Constant | 4.951$^{\bullet\bullet\bullet}$ | 4.906$^{\bullet\bullet\bullet}$ | 5.391$^{\bullet\bullet\bullet}$ | 5.350$^{\bullet\bullet\bullet}$ | 5.531$^{\bullet\bullet\bullet}$ | 5.481$^{\bullet\bullet\bullet}$ |
|  | (0.66) | (0.66) | (0.61) | (0.62) | (0.65) | (0.65) |
| i.demo | Y | Y | Y | Y | Y | Y |
| Observations | 329 | 329 | 329 | 329 | 329 | 329 |
| R-squared | 0.248 | 0.249 | 0.215 | 0.216 | 0.245 | 0.246 |

*Notes. Robust standard errors in parentheses. All p values in this table are two-tailed. "Y" means model includes the fixed effect of assigned demo. "_L" refers to rating scores on lyrics only.*
$^{\bullet\bullet\bullet}$ *p<0.001,* $^{\bullet\bullet}$ *p<0.01,* $^{\bullet}$ *p<0.05,* $^{+}$ *p<0.1*

**Table 5 Regression Results of AI Use on Creativity Measured by Songs (Experiment 2)**

| VARIABLES | (1) Novelty_S | (2) Novelty_S | (3) Usefulness_S | (4) Usefulness_S | (5) Overall Impression_S | (6) Overall Impression_S |
|---|---|---|---|---|---|---|
| AI Use | | 0.133$^+$ | | 0.108$^+$ | | 0.087 |
| | | (0.07) | | (0.06) | | (0.06) |
| Education | 0.143$^{\bullet\bullet}$ | 0.138$^{\bullet\bullet}$ | 0.115$^{\bullet\bullet}$ | 0.111$^{\bullet\bullet}$ | 0.093$^\bullet$ | 0.090$^\bullet$ |
| | (0.05) | (0.05) | (0.04) | (0.04) | (0.04) | (0.04) |
| IQ | 0.005 | 0.005 | 0.006 | 0.006 | 0.007 | 0.007 |
| | (0.01) | (0.01) | (0.01) | (0.01) | (0.01) | (0.01) |
| Specific Human Capital | 0.061 | 0.051 | 0.047 | 0.039 | 0.061 | 0.055 |
| | (0.05) | (0.05) | (0.04) | (0.04) | (0.04) | (0.04) |
| Age | 0.012$^\bullet$ | 0.011$^\bullet$ | 0.003 | 0.003 | -0.002 | -0.002 |
| | (0.01) | (0.01) | (0.01) | (0.01) | (0.00) | (0.00) |
| Gender | 0.046 | 0.034 | 0.062 | 0.052 | 0.039 | 0.031 |
| | (0.07) | (0.07) | (0.06) | (0.06) | (0.06) | (0.06) |
| Openness | -0.085 | -0.072 | -0.041 | -0.031 | -0.060 | -0.052 |
| | (0.06) | (0.06) | (0.05) | (0.05) | (0.05) | (0.05) |
| AI Use frequency | -0.006 | -0.007 | -0.012 | -0.013 | -0.015 | -0.016 |
| | (0.02) | (0.02) | (0.02) | (0.02) | (0.02) | (0.02) |
| Purpose for Experiment | 0.186 | 0.189 | 0.182$^+$ | 0.185$^+$ | 0.229$^\bullet$ | 0.231$^\bullet$ |
| | (0.15) | (0.16) | (0.11) | (0.11) | (0.11) | (0.11) |
| Mind Perception | 0.038 | 0.041 | -0.009 | -0.006 | -0.027 | -0.025 |
| | (0.05) | (0.05) | (0.03) | (0.03) | (0.03) | (0.03) |
| AI Identification Ratio_S | -1.522$^{\bullet\bullet\bullet}$ | -1.620$^{\bullet\bullet\bullet}$ | -1.540$^{\bullet\bullet\bullet}$ | -1.620$^{\bullet\bullet\bullet}$ | -1.300$^{\bullet\bullet\bullet}$ | -1.364$^{\bullet\bullet\bullet}$ |
| | (0.34) | (0.33) | (0.30) | (0.30) | (0.31) | (0.30) |
| Constant | 5.018$^{\bullet\bullet\bullet}$ | 4.986$^{\bullet\bullet\bullet}$ | 5.866$^{\bullet\bullet\bullet}$ | 5.840$^{\bullet\bullet\bullet}$ | 6.020$^{\bullet\bullet\bullet}$ | 5.999$^{\bullet\bullet\bullet}$ |
| | (0.48) | (0.48) | (0.37) | (0.37) | (0.35) | (0.36) |
| i.demo | Y | Y | Y | Y | Y | Y |
| Observations | 329 | 329 | 329 | 329 | 329 | 329 |
| R-squared | 0.440 | 0.446 | 0.448 | 0.454 | 0.508 | 0.512 |

*Notes. Robust standard errors in parentheses. All p values in this table are two-tailed. "Y" means model includes the fixed effect of assigned demo. "_S" refers to rating scores with songs hearing.*
$^{\bullet\bullet\bullet}$ *p<0.001,* $^{\bullet\bullet}$ *p<0.01,* $^\bullet$ *p<0.05,* $^+$ *p<0.1*

**Table 6 Regression Results of Interaction of AI Use and Education on Creativity (Experiment 2)**

| VARIABLES | (1) Novelty_L | (2) Usefulness_L | (3) Overall Impression_L | (4) Novelty_S | (5) Usefulness_S | (6) Overall Impression_S |
|---|---|---|---|---|---|---|
| AI Use | 0.100 | 0.090 | 0.113 | 0.137$^{+}$ | 0.113$^{+}$ | 0.092 |
| | (0.11) | (0.11) | (0.12) | (0.07) | (0.06) | (0.06) |
| Education | -0.106 | -0.036 | -0.117 | 0.043 | -0.012 | -0.013 |
| | (0.12) | (0.13) | (0.13) | (0.10) | (0.09) | (0.09) |
| AI Use × Education | 0.407$^{••}$ | 0.341$^{•}$ | 0.489$^{••}$ | 0.146 | 0.192$^{+}$ | 0.159$^{+}$ |
| | (0.15) | (0.15) | (0.16) | (0.11) | (0.10) | (0.10) |
| IQ | 0.028 | 0.015 | 0.013 | 0.004 | 0.005 | 0.007 |
| | (0.02) | (0.02) | (0.02) | (0.01) | (0.01) | (0.01) |
| Specific Human Capital | 0.094 | 0.098 | 0.165$^{•}$ | 0.046 | 0.032 | 0.049 |
| | (0.09) | (0.08) | (0.08) | (0.05) | (0.04) | (0.04) |
| Age | 0.011 | 0.015 | 0.008 | 0.011$^{•}$ | 0.002 | -0.002 |
| | (0.01) | (0.01) | (0.01) | (0.01) | (0.01) | (0.00) |
| Gender | 0.116 | 0.185$^{+}$ | 0.185 | 0.034 | 0.052 | 0.031 |
| | (0.11) | (0.11) | (0.12) | (0.07) | (0.06) | (0.06) |
| Openness | -0.086 | -0.112 | -0.150 | -0.073 | -0.031 | -0.052 |
| | (0.09) | (0.09) | (0.09) | (0.06) | (0.05) | (0.05) |
| AI Use frequency | -0.027 | -0.001 | -0.011 | -0.006 | -0.013 | -0.015 |
| | (0.04) | (0.04) | (0.03) | (0.02) | (0.02) | (0.02) |
| Purpose for Experiment | 0.083 | -0.108 | -0.008 | 0.190 | 0.186$^{+}$ | 0.232$^{•}$ |
| | (0.24) | (0.23) | (0.24) | (0.16) | (0.11) | (0.11) |
| Mind Perception | -0.090 | -0.086 | -0.108$^{+}$ | 0.037 | -0.012 | -0.030 |
| | (0.07) | (0.06) | (0.06) | (0.05) | (0.03) | (0.03) |
| AI Identification Ratio_L | -1.434$^{•••}$ | -1.423$^{•••}$ | -1.294$^{•••}$ | | | |
| | (0.21) | (0.21) | (0.22) | | | |
| AI Identification Ratio_S | | | | -1.623$^{•••}$ | -1.624$^{•••}$ | -1.368$^{•••}$ |
| | | | | (0.33) | (0.29) | (0.29) |
| Constant | 5.692$^{•••}$ | 6.234$^{•••}$ | 6.470$^{•••}$ | 5.612$^{•••}$ | 6.364$^{•••}$ | 6.422$^{•••}$ |
| | (0.63) | (0.62) | (0.66) | (0.44) | (0.36) | (0.35) |
| | | | | | | |
| i.demo | Y | Y | Y | Y | Y | Y |
| Observations | 329 | 329 | 329 | 329 | 329 | 329 |
| R-squared | 0.267 | 0.230 | 0.271 | 0.450 | 0.464 | 0.518 |

*Notes. Robust standard errors in parentheses. All p values in this table are two-tailed. "Y" means model includes the fixed effect of assigned demo. "_L" refers to rating scores on lyrics only; "_S" refers to rating scores with songs hearing.*
$^{•••}$ *p<0.001,* $^{••}$ *p<0.01,* $^{•}$ *p<0.05,* $^{+}$ *p<0.1*

**Table 7 Regression Results of Interaction of AI Use and IQ on Creativity (Experiment 2)**

| VARIABLES | (1)<br>Novelty_L | (2)<br>Usefulness_L | (3)<br>Overall<br>Impression_L | (4)<br>Novelty_S | (5)<br>Usefulness_S | (6)<br>Overall<br>Impression_S |
|---|---|---|---|---|---|---|
| AI Use | 0.083 | 0.077 | 0.095 | 0.132[+] | 0.108[+] | 0.088 |
| | (0.11) | (0.11) | (0.12) | (0.07) | (0.06) | (0.06) |
| IQ | -0.028 | -0.015 | -0.022 | -0.007 | 0.005 | 0.016 |
| | (0.04) | (0.04) | (0.04) | (0.02) | (0.02) | (0.02) |
| AI Use × IQ | 0.076[+] | 0.042 | 0.050 | 0.016 | 0.002 | -0.011 |
| | (0.04) | (0.04) | (0.04) | (0.02) | (0.02) | (0.02) |
| Education | 0.160• | 0.186• | 0.201• | 0.138•• | 0.111•• | 0.089• |
| | (0.08) | (0.08) | (0.08) | (0.05) | (0.04) | (0.04) |
| Specific Human Capital | 0.109 | 0.111 | 0.184• | 0.051 | 0.039 | 0.055 |
| | (0.08) | (0.08) | (0.08) | (0.05) | (0.04) | (0.04) |
| Age | 0.011 | 0.015 | 0.008 | 0.011• | 0.003 | -0.001 |
| | (0.01) | (0.01) | (0.01) | (0.01) | (0.01) | (0.00) |
| Gender | 0.116 | 0.185[+] | 0.185 | 0.033 | 0.052 | 0.032 |
| | (0.11) | (0.11) | (0.12) | (0.07) | (0.06) | (0.06) |
| Openness | -0.106 | -0.122 | -0.162[+] | -0.077 | -0.031 | -0.049 |
| | (0.09) | (0.09) | (0.09) | (0.06) | (0.05) | (0.05) |
| AI Use frequency | -0.032 | -0.004 | -0.015 | -0.008 | -0.013 | -0.015 |
| | (0.04) | (0.04) | (0.03) | (0.02) | (0.02) | (0.02) |
| Purpose for Experiment | 0.112 | -0.093 | 0.008 | 0.196 | 0.186[+] | 0.227• |
| | (0.25) | (0.23) | (0.25) | (0.15) | (0.11) | (0.11) |
| Mind Perception | -0.096 | -0.086 | -0.106[+] | 0.037 | -0.007 | -0.023 |
| | (0.07) | (0.06) | (0.06) | (0.05) | (0.03) | (0.03) |
| AI Identification Ratio_L | -1.405••• | -1.398••• | -1.259••• | | | |
| | (0.22) | (0.21) | (0.22) | | | |
| AI Identification Ratio_S | | | | -1.625••• | -1.621••• | -1.361••• |
| | | | | (0.33) | (0.30) | (0.30) |
| Constant | 5.549••• | 5.712••• | 5.843••• | 5.110••• | 5.935••• | 6.073••• |
| | (0.64) | (0.60) | (0.64) | (0.48) | (0.36) | (0.37) |
| i.demo | Y | Y | Y | Y | Y | Y |
| Observations | 329 | 329 | 329 | 329 | 329 | 329 |
| R-squared | 0.258 | 0.219 | 0.250 | 0.447 | 0.454 | 0.512 |

*Notes. Robust standard errors in parentheses. All p values in this table are two-tailed. "Y" means model includes the fixed effect of assigned demo. "_L" refers to rating scores on lyrics only; "_S" refers to rating scores with songs hearing.*
*••• p<0.001, •• p<0.01, • p<0.05, [+] p<0.1*

**Table 8 Regression Results of Interaction of AI Use and Specific Human Capital on Creativity**

**(Experiment 2)**

| VARIABLES | (1) Novelty_L | (2) Usefulness_L | (3) Overall Impression_L | (4) Novelty_S | (5) Usefulness_S | (6) Overall Impression_S |
|---|---|---|---|---|---|---|
| AI Use | 0.074 | 0.068 | 0.088 | 0.124$^+$ | 0.100$^+$ | 0.079 |
| | (0.11) | (0.11) | (0.12) | (0.07) | (0.06) | (0.06) |
| Specific Human Capital | 0.377$^{\bullet\bullet}$ | 0.334$^\bullet$ | 0.378$^{\bullet\bullet\bullet}$ | 0.204$^\bullet$ | 0.174$^\bullet$ | 0.198$^{\bullet\bullet}$ |
| | (0.14) | (0.15) | (0.14) | (0.09) | (0.07) | (0.07) |
| AI Use × Specific Human Capital | -0.391$^\bullet$ | -0.327$^+$ | -0.283$^+$ | -0.221$^\bullet$ | -0.195$^\bullet$ | -0.209$^\bullet$ |
| | (0.16) | (0.17) | (0.17) | (0.10) | (0.08) | (0.08) |
| Education | 0.168$^\bullet$ | 0.193$^\bullet$ | 0.207$^\bullet$ | 0.144$^{\bullet\bullet}$ | 0.117$^{\bullet\bullet}$ | 0.096$^\bullet$ |
| | (0.08) | (0.08) | (0.08) | (0.05) | (0.04) | (0.04) |
| IQ | 0.031$^+$ | 0.018 | 0.016 | 0.006 | 0.007 | 0.008 |
| | (0.02) | (0.02) | (0.02) | (0.01) | (0.01) | (0.01) |
| Age | 0.013 | 0.016$^+$ | 0.010 | 0.012$^\bullet$ | 0.003 | -0.002 |
| | (0.01) | (0.01) | (0.01) | (0.00) | (0.01) | (0.00) |
| Gender | 0.124 | 0.192$^+$ | 0.191 | 0.039 | 0.057 | 0.036 |
| | (0.11) | (0.11) | (0.12) | (0.07) | (0.06) | (0.06) |
| Openness | -0.059 | -0.090 | -0.130 | -0.058 | -0.018 | -0.038 |
| | (0.10) | (0.09) | (0.09) | (0.06) | (0.05) | (0.05) |
| AI Use frequency | -0.022 | 0.004 | -0.008 | -0.003 | -0.010 | -0.012 |
| | (0.04) | (0.03) | (0.03) | (0.02) | (0.02) | (0.02) |
| Purpose for Experiment | 0.085 | -0.106 | -0.009 | 0.193 | 0.188$^+$ | 0.235$^\bullet$ |
| | (0.24) | (0.23) | (0.24) | (0.15) | (0.11) | (0.11) |
| Mind Perception | -0.085 | -0.082 | -0.099 | 0.036 | -0.011 | -0.030 |
| | (0.07) | (0.06) | (0.06) | (0.05) | (0.03) | (0.03) |
| AI Identification Ratio_L | -1.391$^{\bullet\bullet\bullet}$ | -1.387$^{\bullet\bullet\bullet}$ | -1.249$^{\bullet\bullet\bullet}$ | | | |
| | (0.22) | (0.21) | (0.23) | | | |
| AI Identification Ratio_S | | | | -1.580$^{\bullet\bullet\bullet}$ | -1.585$^{\bullet\bullet\bullet}$ | -1.327$^{\bullet\bullet\bullet}$ |
| | | | | (0.33) | (0.30) | (0.30) |
| Constant | 4.907$^{\bullet\bullet\bullet}$ | 5.379$^{\bullet\bullet\bullet}$ | 5.633$^{\bullet\bullet\bullet}$ | 4.953$^{\bullet\bullet\bullet}$ | 5.802$^{\bullet\bullet\bullet}$ | 5.977$^{\bullet\bullet\bullet}$ |
| | (0.63) | (0.60) | (0.62) | (0.47) | (0.36) | (0.34) |
| i.demo | Y | Y | Y | Y | Y | Y |
| Observations | 329 | 329 | 329 | 329 | 329 | 329 |
| R-squared | 0.261 | 0.226 | 0.253 | 0.453 | 0.461 | 0.519 |

*Notes. Robust standard errors in parentheses. All p values in this table are two-tailed. "Y" means model includes the fixed effect of assigned demo. "_L" refers to rating scores on lyrics only; "_S" refers to rating scores with songs hearing.*
$^{\bullet\bullet\bullet}$ *p<0.001,* $^{\bullet\bullet}$ *p<0.01,* $^\bullet$ *p<0.05,* $^+$ *p<0.1*

**Figure 1 Design of Experiment 1[3]**

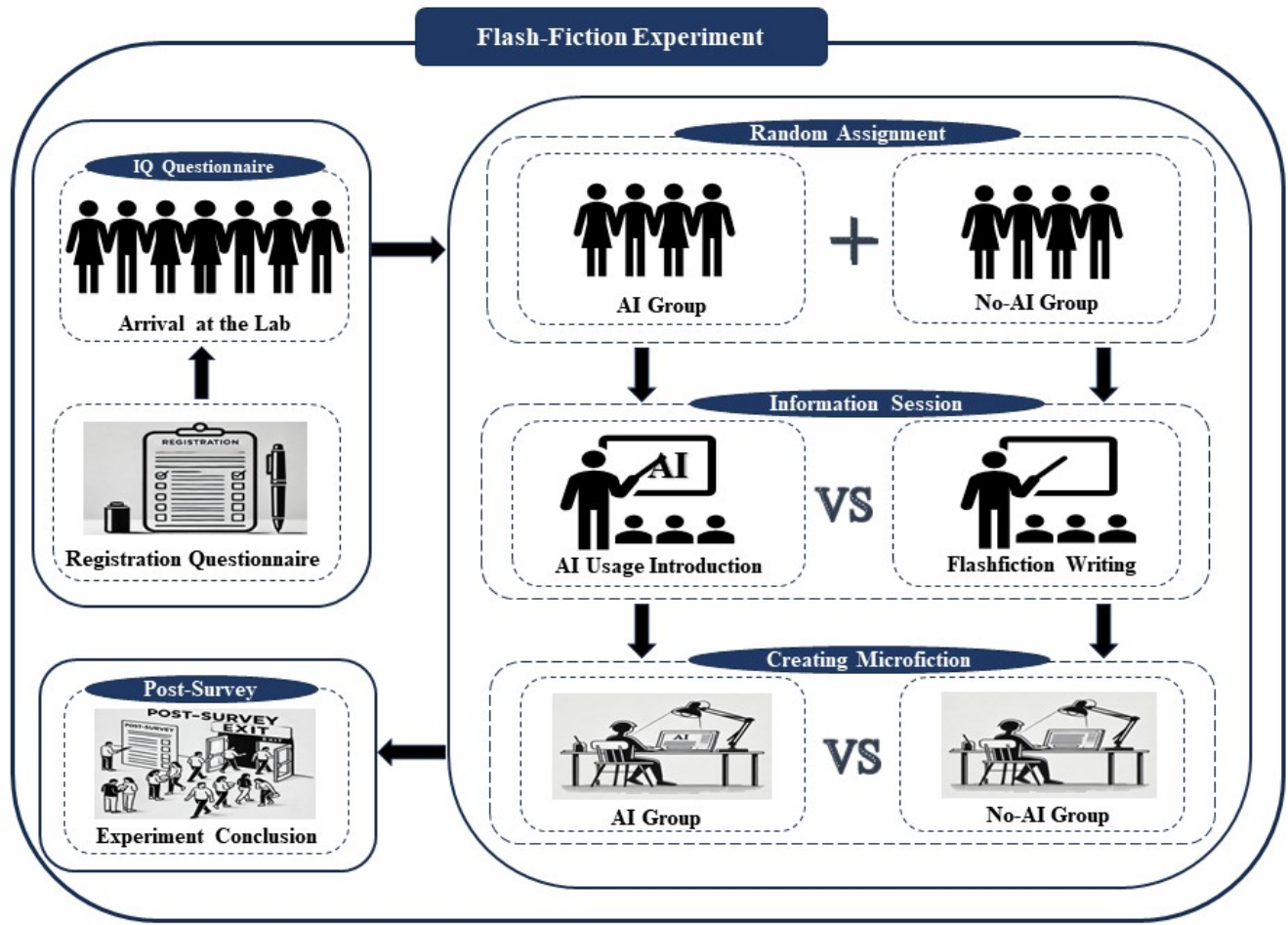

---

[3] Figures and icons are generated by AI.

**Figure 2 Design of Experiment 2[4]**

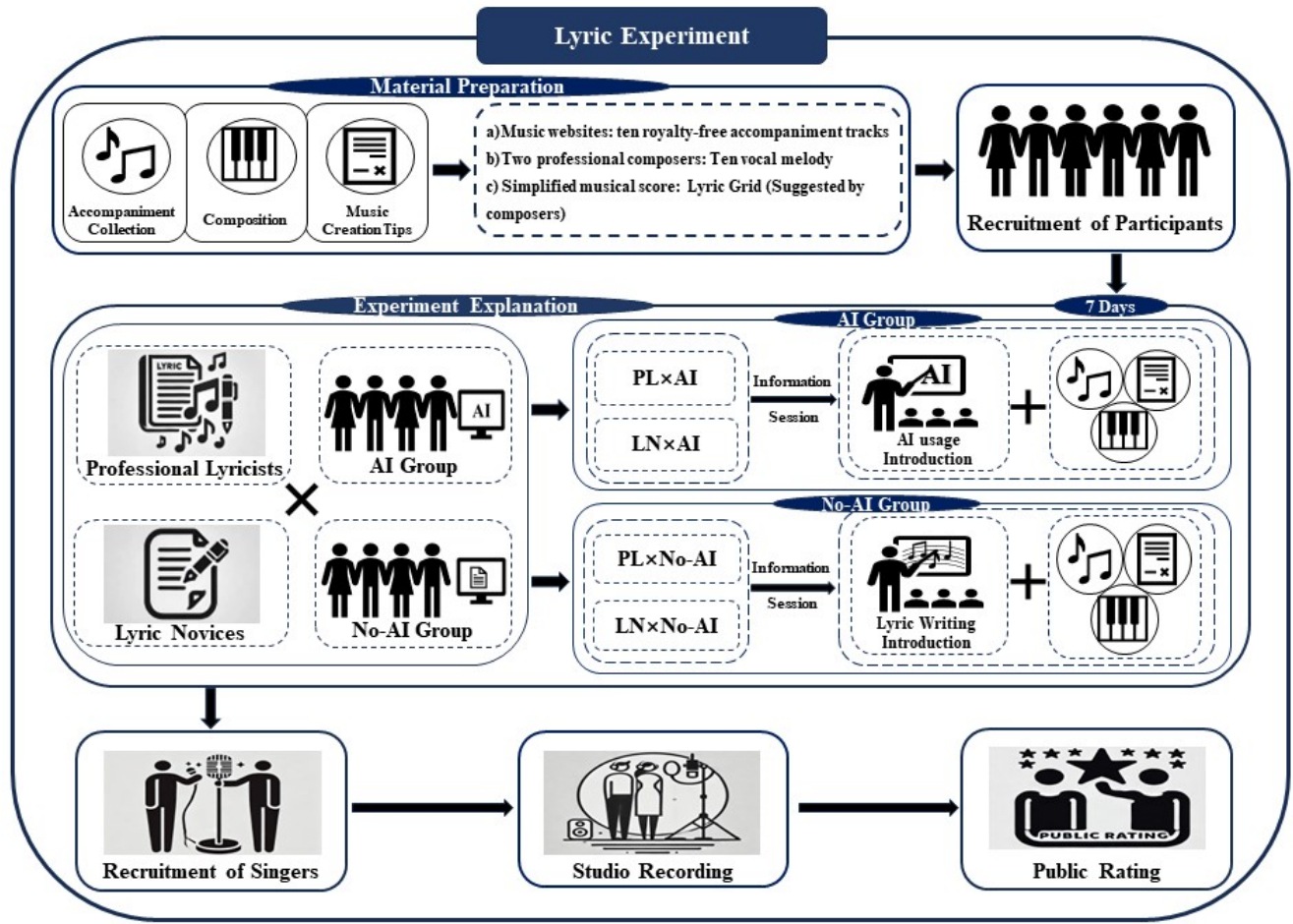

---

[4] Figures and icons are generated by AI. "PL" stands for Professional Lyricists, and "LN" stands for Lyric Novices.

**Figure 3 AI Tool Used in Two Experiments**

**Figure 4 Interaction of AI Use and Education Predicting Novelty Score**

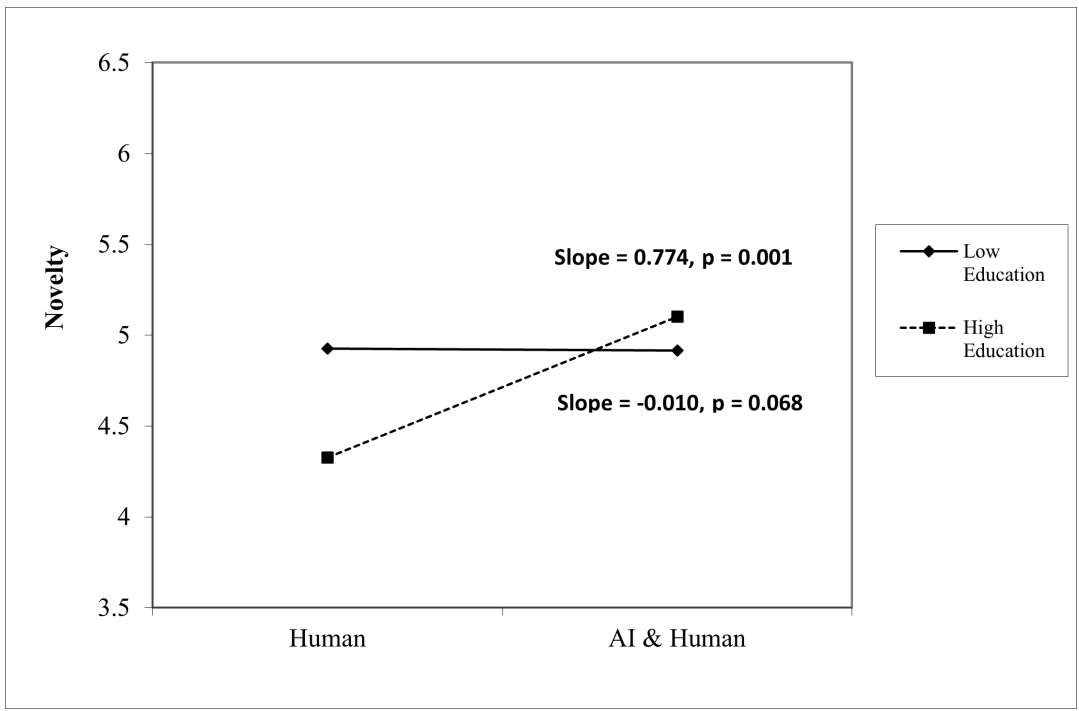

**Figure 5 Interaction of AI Use and IQ Predicting Novelty Score**

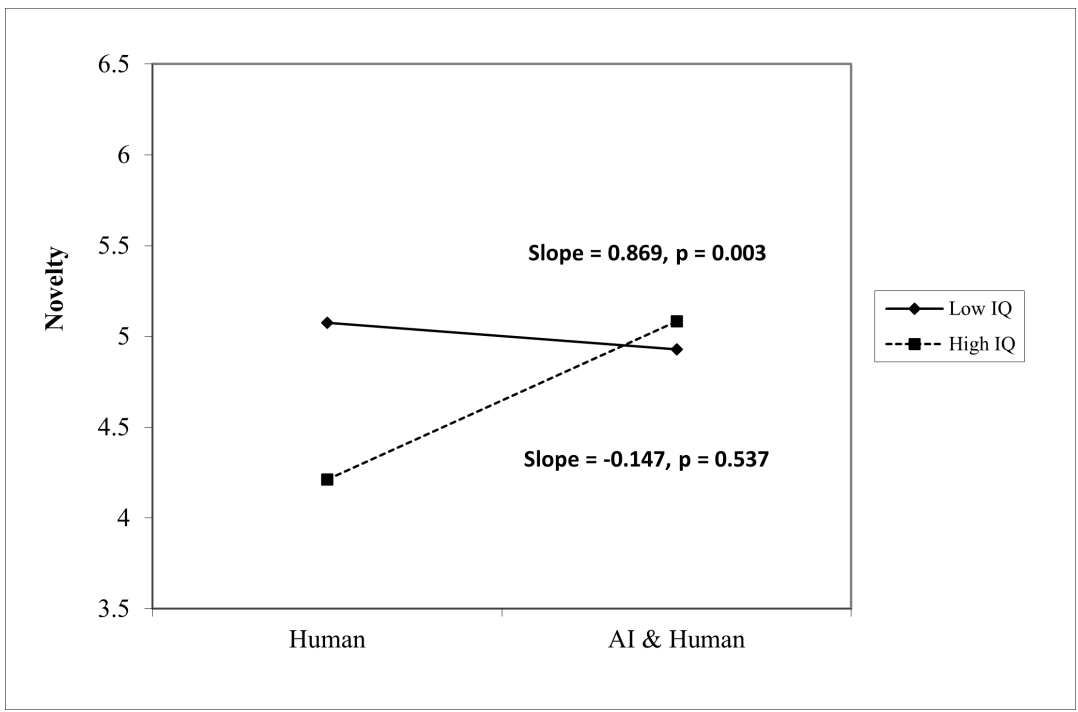

**Figure 6 Interaction of AI Use and Specific Human Capital Predicting Novelty Score**

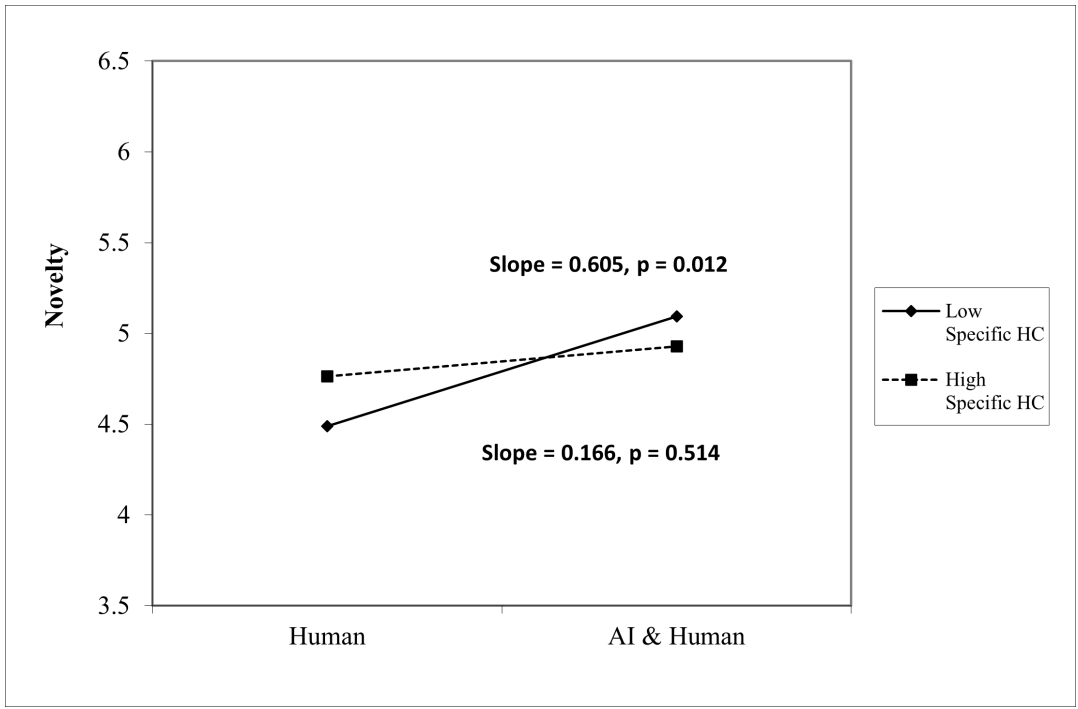

**Figure 7 Interaction of AI Use and Education Predicting Novelty Score by Song Rating**

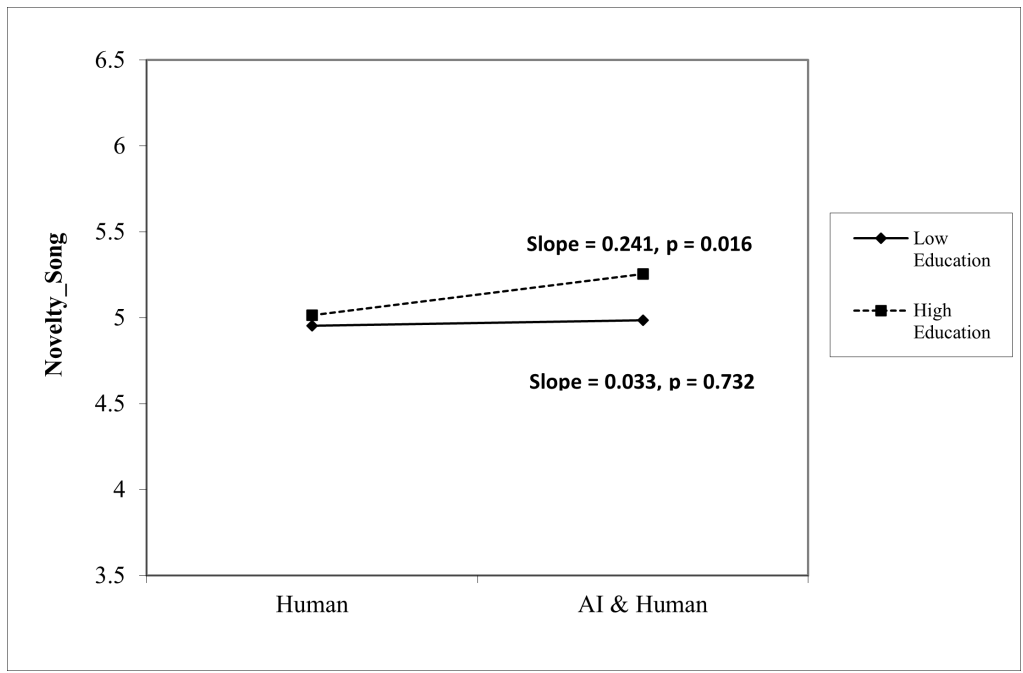

**Figure 8 Interaction of AI Use and Specific Human Capital Predicting Novelty Score by**

**Song Rating**

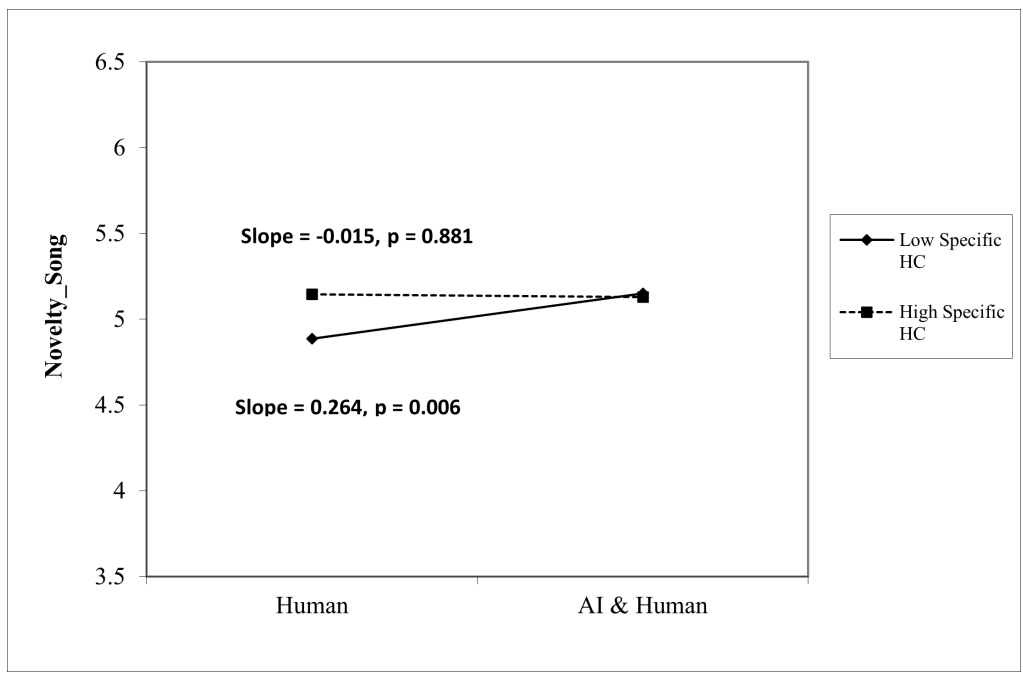